\documentclass[usenatbib]{mnras}

\usepackage[T1]{fontenc}
\usepackage{ae,aecompl}

\usepackage{graphicx}	

\usepackage{amsmath}	

\usepackage{amssymb}	

\usepackage{deluxetable}

\usepackage{amsmath,amssymb,amsfonts}
\usepackage{color}
\usepackage{arydshln}
\usepackage{epsfig}

\newcommand \beq {\begin{equation}}
\newcommand \enq {\end{equation}}

\newcommand \reynolds {\rho v_{x} \delta v_{y}}

\definecolor{grey}{gray}{0.5}

\newcommand \lx  {L_{\rm {x}}}
\newcommand \ly  {L_{\rm {y}}}
\newcommand \lz  {L_{\rm {z}}}
\newcommand \Bx  {B_{\rm {x}}}
\newcommand \By  {B_{\rm {y}}}

\newcommand \bx  {b_{\rm {x}}}
\newcommand \by  {b_{\rm {y}}}

\newcommand \cs  {c_{\rm {s}}}
\newcommand \ar  {\alpha_{\rm R}}
\newcommand \am  {\alpha_{\rm M}}
\newcommand \Langle {\left\langle}
\newcommand \Rangle {\right\rangle}
\newcommand \Tcycle {T_{\rm {cycle}}}
\newcommand \dLangle  {\Langle\!\Langle}
\newcommand \dRangle  {\Rangle\!\Rangle}

\newcommand \ob       {\overline}
\newcommand \wt       {\widetilde}
\newcommand \Bxk     {\ob{B}^{(1)}_x} 
\newcommand \Byk     {\ob{B}^{(1)}_y}

\title[MRI in unstratified zero-net-flux shearing box]{Saturation of the magnetorotational instability in the unstratified shearing box 
with zero net flux: convergence in taller boxes}
\author[J.-M. Shi et al.]{
Ji-Ming Shi,$^{1}$\thanks{E-mail:~jmshi@astro.princeton.edu}
James M. Stone,$^{1}$
Chelsea X. Huang $^{1}$
\\
$^{1}$ {Department of Astrophysical Sciences, Princeton University, 4 Ivy Ln,
    Princeton, NJ 08544}
}

\date{Accepted XXX. Received YYY; in original form ZZZ}

\pubyear{2015}

\begin{document}
\label{firstpage}
\pagerange{\pageref{firstpage}--\pageref{lastpage}}
\maketitle

\begin{abstract}
Previous studies of the nonlinear regime of the magnetorotational
instability in one particular type of shearing box model --
unstratified with no net magnetic flux -- find that without explicit
dissipation (viscosity and resistivity) the saturation amplitude
decreases with increasing numerical resolution.
We show that this result is strongly dependent on the vertical aspect ratio
of the computational domain
$\lz/\lx$.  When $\lz/\lx\lesssim 1$, we
recover previous results.  However, when the vertical domain is
extended $\lz/\lx \gtrsim 2.5$, we find the saturation level
of the stress is greatly increased (giving a ratio of stress to
pressure $\alpha \gtrsim 0.1$), and moreover the results are independent
of numerical resolution.  Consistent with previous results, we find
that saturation of the MRI in this
regime is controlled by a cyclic dynamo which generates patches of
strong toroidal field that switches sign on scales of
$\lx$ in the vertical direction.  We speculate that when $\lz/\lx\lesssim 1$, the dynamo is inhibited by
the small size of the vertical domain, leading to the puzzling dependence of
saturation amplitude on resolution.
We show that previous toy
models developed to explain the MRI dynamo are
consistent with our results, and that the cyclic pattern of toroidal
fields observed in stratified shearing box simulations (leading to the so-called
butterfly diagram) may also be related.  
In tall boxes the saturation amplitude is insensitive
to whether or not explicit
dissipation is included in the calculations, at least for large magnetic
Reynolds and Prandtl number. 
Finally, we show MRI turbulence in tall domains has a smaller critical $\rm{Pm}_c$, 
and an extended lifetime compared to $\lz/\lx\lesssim 1$ boxes.
\end{abstract}

\begin{keywords}
{accretion, accretion disks -- dynamo -- magnetohydrodynamics (MHD) 
-- turbulence -- instabilities -- methods: numerical  }
\end{keywords}

\section{INTRODUCTION}
\label{sec:introduction}
The local shearing box model has proved to be very useful for studying the non-linear
regime of the magneto-rotational instability (MRI)\citep[][]{hgb95}. 
It is useful to classify such models into four types, depending on whether
or not
the domain contains net magnetic flux, and whether or not the vertical component of
gravity is included (producing a vertically stratified density profile). Generally speaking, the
results of numerical simulations that explore three of these four types of shearing box models can
be summarized as follows.
\begin{enumerate}
\item Unstratified shearing box simulations with net flux show sustained MHD turbulence and 
numerically converged values of the stress
and angular momentum transport.  The saturated stress-to-pressure ratio $\alpha$
varies widely ($\sim 10^{-3}$-$0.1$) depending on the net magnetic field strength
\citep[][]{hgb95,sano2004,Guan2009,simonetal2009}.
\item Vertically stratified shearing box simulations with no net flux also show strong MRI
	driven turbulence with typical $\alpha\sim 10^{-3}$-$10^{-2}$, as well as
dynamo activity that leads to a quasi-periodic pattern of alternating
toroidal field \citep[][]{Brandenburgetal1995,shgb96,Gressel2010,Simon13a}.  The
	stress is independent of numerical resolution \citep[][]{Shi2010,davisetal2010}, 
	although recently \citet[][]{Bodoetal2014} have claimed at very high resolution
this may no longer be the case.  Further investigation of this issue is required.
\item Vertically stratified shearing box simulations with net flux show a wide range
	of behavior depending on the field geometry and strength \citep[][]{shgb96,MS2000}.  
For example, sustained
turbulence is observed with weak toroidal fields, while powerful outflows that depend
on the field strength are produced in the
case of net vertical fields, and the disk can show complex interplay between MRI and
buoyancy (Parker) instabilities for strong toroidal fields
\citep[][]{suzuki2009,GG2011,SHB2011,FLLO2013,LFO2013,bai_stone2013,Simon13b}.
\end{enumerate}

The final case of
an unstratified shearing box with no magnetic flux, introduced by \citet[][]{hgb96} to
study the MHD dynamo driven by the MRI,
shows intriguing behavior.  In this case simulations find that the saturated level
of stress decreases
as the numerical resolution is
increased\citep[][]{FP2007,simonetal2009,Guan2009,bodoetal2011}. 
When physical dissipation (viscosity and resistivity) is included in the model, convergence
of the stress with resolution is recovered\citep{FPLH2007}. 
Numerous works have also shown that the ratio of these two dissipation length scales 
seems to control the saturation level of stress driven by MRI turbulence
\citep[][]{FPLH2007,simonetal2009,Fromang2010}, even when net flux is included
\citep[][]{LL2007,SH2009,Meheutetal2015}. 

Since the important work of \citet{FP2007} there has been significant effort
to investigate MHD dynamo action in the special case of unstratified no-net flux shearing
box.  
Almost all of this work is based on spectral methods, so that explicit 
dissipation (both viscosity and resistivity) must be included. 
For example, \citet{LO08} found
that long-lived cyclic dynamo activity occurs in an
incompressible MHD model in which the aspect ratio of the computational domain is
$(\lx : \ly : \lz) = (1 : 4 : 2)$,
and these authors developed a toy dynamical model to explain the cycles.  More recently,
the work of \citet{Rincon2007, Rincon2008} has explored the subcritical dynamo mechanism in
more detail, and in particular the role of the magnetic Prandtl number (ratio of viscosity to
resistivity) in determining the outcome.  Recent work \citep[][]{Riolsetal2013, Riolsetal2015}  
reveals the nonlinear dynamics of the MRI dynamo is more complex than expected.

Despite this progress, the question remains why, in compressible MHD with
no explicit dissipation,
does the saturation amplitude of the MRI depend so strongly on resolution?
One hint may come from the extended vertical domain used in the study of MRI dynamos
in incompressible MHD, e.g. \citet{LO08}.  Most previous studies in
compressible MHD adopt a standard computational domain that spans one thermal scale
height $H$ in the radial direction, and uses a vertical aspect ratio $\lz/\lx=1$ and
a toroidal aspect ratio $\ly/\lx \gtrsim 4$.  In contrast, studies of shear-driven
dynamos have found that a large vertical aspect ratio is required for vigorous
dynamo action \citep{Yousefetal2008}, consistent with the results of \citet{LO08}.

In this work, we present new studies of the saturation of the MRI in compressible
MHD in the no-net flux unstratified shearing box, both with and without explicit dissipation,
to investigate the role of the size and aspect ratio of the computational domain
on the results.  Similar to the results reported in incompressible MHD, we
find qualitatively different behavior when the vertical aspect ratio of the domain
is large, $\lz/\lx \gtrsim 2.5$.  In such case vigorous dynamo action produces a strong,
cyclic toroidal magnetic field that greatly increases the level of turbulence and stress.
Moreover the saturation amplitude of the turbulence is independent of numerical resolution,
and independent of whether explicit dissipation is included in the model (at least for
large magnetic Reynolds numbers and magnetic Prandtl numbers greater than one).

In reality, the unstratified shearing box with no net flux has very
little relevance for real astrophysical disks, because it is
impossible for each local patch in a global disk to maintain
zero-net-flux for all time.  Even if overall the disk has no net
flux, large scale field loops on the scale of the vertical size of
the disk will impart smaller local patches with a time-varying net
flux.  Moreover, in many astrophysical plasmas the dominant non-ideal
MHD effect is not simply Ohmic dissipation, but rather ambipolar
diffusion and/or the Hall effect \citep[][]{Wardle1999,SW2003}.
Significant progress has been made recently on investigating the
role of ambipolar diffusion
\citep[][]{HS1998,BS2011,BS2013,Simon13a,Simon13b} and the Hall
effect \citep[][]{sano2002,kunz13,lesur14,Bai2014,Bai2015} on the
saturation of the MRI in all four kinds of shearing box models
described earlier, and the results differ significantly from models
which include only Ohmic resistivity.  Nevertheless, it is of
interest to understand dynamo action in the unstratified no net
flux shearing box (with and without resistivity) if only because
it represents such a simple well-posed model.  In addition, we show
in this paper that the cyclic dynamo action observed in {\em
stratified} shearing box models (resulting in the so-called butterfly
diagram of toroidal field) may be related to that observed in
more realistic unstratified domains.

The structure of the paper is as follows: we first describe the
equations we solve and the numerical methods adopted in
section~\ref{sec:methods}; the main results are then presented in
section~\ref{sec:results}; a short discussion of the dynamo action
observed in our results follows in section~\ref{sec:dynamo}; and
finally in section~\ref{sec:conclusion} we summarize and conclude.

\section{METHODS}
\label{sec:methods}

\subsection{Equations solved and code description \label{sec:eqn}}
We solve the comprssible MHD equations adopting
the ``shearing box'' approximation. In a Cartesian reference frame corotating with the disk at fixed
orbital frequency $\Omega \hat{\mathbf{z}}$, the equations solved are as follows:
\begin{align}
  \frac{\partial \rho}{\partial t} + \nabla\cdot (\rho \mathbf{v}) = 0 \,, 
  \label{eq:continuity} \\
  \frac{\partial \rho\mathbf{v}}{\partial t} + \nabla\cdot\left(\rho\mathbf{v}\mathbf{v}
  + \mathbf{T} \right) =
  - 2\rho \Omega \hat{\mathbf{z}} \times\mathbf{v}
  +2q\rho\Omega^2 x \hat{\mathbf{x}} \,, 
  \label{eq:eom} \\
  \frac{\partial \mathbf{B}}{\partial t} - \nabla \times \left(\mathbf{v}\times\mathbf{B} \,
  - \,\eta\nabla\times \mathbf{B}\right) = 0 \,, 
  \label{eq:induction}
\end{align}
where $\hat{\mathbf{x}}$ refers to the radial direction,
$\rho$ is the mass density, $\mathbf{v}$ is the velocity, $q = 3/2$ is the Keplerian
shear parameter, $\mathbf{B}$ is the magnetic strength, and $\eta$ is the Ohmic resistivity. 
The total stress tensor $\mathbf{T}$ is defined as
\beq
\mathbf{T} = \left(P+\frac{\mathbf{B}\cdot\mathbf{B}}{8\pi} \right)\mathbf{I} -
\frac{\mathbf{B}\mathbf{B}}{4\pi}
  \, - \,\mathbf{\Pi}
  \,,
  \label{eq:tensor}
\enq
where $\mathbf{I}$ is the identity tensor, $P=\rho \cs^2$ is the gas pressure, and $\cs$ is the
isothermal sound speed. 
The viscous stress $\mathbf{\Pi}$ can be expanded as 
\beq
\Pi_{ij} = \rho\nu\left(\frac{\partial v_i}{\partial x_j}+\frac{\partial v_j}{\partial x_j}
-\frac{2}{3}\delta_{ij}\nabla\cdot\mathbf{v}\right) \,,
\enq
where $\nu$ is the kinematic viscosity.
Note that in Equation~(\ref{eq:eom}), the vertical tidal acceleration $\Omega^2
z\hat{\mathbf{z}}$ is omitted, that is we are studying the unstratified shearing box.
We include explicit viscosity and resistivity terms in some of our calculations, as
discussed in section~\ref{sec:diss}.  However, in order to investigate the convergence of
stress with numerical resolution without dissipation, most of simulations
are ideal MHD.

We use the {\texttt{Athena}} \citep{stoneetal08,sg10} MHD code for our numerical simulations.
We adopt the CTU integrator, third-order piecewise parabolic
reconstruction with characteristic tracing in the primitive variables, 
and the Roe Riemann solver as the basic algorithms. 
The orbital advection scheme is used to
increase the efficiency of calculation \citep{Masset2000,Johnson2008}.
As the shearing box
boundary conditions can cause mismatch of the integral of fluxes over the two radial faces due to
remap \citep{GZ2007}, 
using orbital advection can reduce this mismatch and
improve the conservation \citep{sg10}. We ensure the conservation of vertical magnetic flux to 
machine precision by treating the shearing box boundary conditions carefully with a remapping
method described in \citet[][Section~4]{sg10}.

\subsection{Initial conditions and run setup}
\label{sec:ic}

We initialize the disk density with a uniform distribution $\rho=\rho_0=1$ within the box. We set the
unit of time $\Omega^{-1}=1$ and the unit of length $H=\cs/\Omega=1$, and therefore the sound speed $\cs$ and
initial gas pressure $P_0=\rho_0 \cs^2$ are unity as well. The initial magnetic field is
$\mathbf{B}=B_0\sin(2\pi x/\lx)\hat{\mathbf{z}}$, where $B_0$ is specified through the
plasma $\beta_0\equiv P_0/(B_0^2/8\pi)=100$ for all runs.  With this configuration, the
net magnetic flux is zero in all three directions. 
We also ran several models in which the initial magnetic field geometry and strength were
varied to check that our results do not depend on the precise form of the initial
conditions, provided the net flux is zero.
We adopt periodic boundary conditions in the azimuthal ($y$) and vertical ($z$) directions, and 
shearing periodic in the radial ($x$) direction \citep{hgb95}.

We carry out multiple sets of simulations with various box sizes $\lx \in \{0.25,0.5,1.0\}\,H$,
vertical to radial aspect ratios $\lz/\lx = 0.5$-$12$ and numerical resolutions $(\Delta
x)^{-1}=(\Delta y)^{-1} = (\Delta z)^{-1} \in \{32,64,128,256\}\,H^{-1}$. 
Parameters for all our runs are all listed in Table~\ref{tab:tab1}. We do not vary the
toroidal aspect ratio $\ly/\lx$ in 
this work, instead we fix $\ly/\lx = 4$ for all runs.  Provided $\ly/\lx \gtrsim 1$,
it has been found that different
sized domains in the toroidal direction primarily affect the spectrum and amplitude
of spiral density waves excited by the turbulence and amplified by shear \citet[][]{HP2009}, 
however such waves do not dominate the saturation level of stress.

\subsection{Diagnostics}
\label{sec:diag}
In order to facilitate analysis and obtain statistical properties of our simulations, we define a few
ways to average physical variables. 
We first define a volume (box) average:
\beq
\langle X \rangle \equiv \frac{\int \,X \,dx dy dz}{\int dx dy dz} \,.
\label{eq:volavg}
\enq
We also define a time average
\beq
\langle X \rangle_t \equiv \frac{\int \,X \,dt}{\int dt} \,.
\label{eq:timeavg}
\enq
The time average is generally applied over $\sim 200$ orbits to eliminate chaotic
fluctuations and achieve
meaningful statistics \citep[][]{WBH2003}.
Using these averages, we write the time averaged Maxwell and Reynolds stresses as 
\beq
\am \equiv \langle\langle -\frac{\Bx\By}{4\pi P_0}\rangle\rangle_{t} \,, ~
\ar \equiv \langle\langle \frac{\reynolds}{P_0}\rangle\rangle_{t} 
\label{eq:alpha_def}
\enq
respectively. In Table~\ref{tab:tab1} we also list the total internal stress,
$\alpha_{\rm tot}\equiv \ar + \am$.  We use these stresses as a measure of
the saturated level of MRI turbulence throughout this paper. 

As some of our simulations develop large scale coherent field structures within the box, we
therefore introduce a horizontal average:
\beq
\ob{X} \equiv \frac{\int\,X\, dx dy}{\int dx dy} \,.
\label{eq:havg}
\enq
This average is useful for decomposing the total magnetic field $\mathbf{B}$
into mean and fluctuating parts. i.e.
\beq
\mathbf{B} = \ob{\mathbf{B}} + \mathbf{b} \,,
\enq
where $\mathbf{b}$ denotes the turbulent (small scale) fluctuating field.

\section{RESULTS}
\label{sec:results}
\begin{figure*}
\includegraphics[width=16cm]{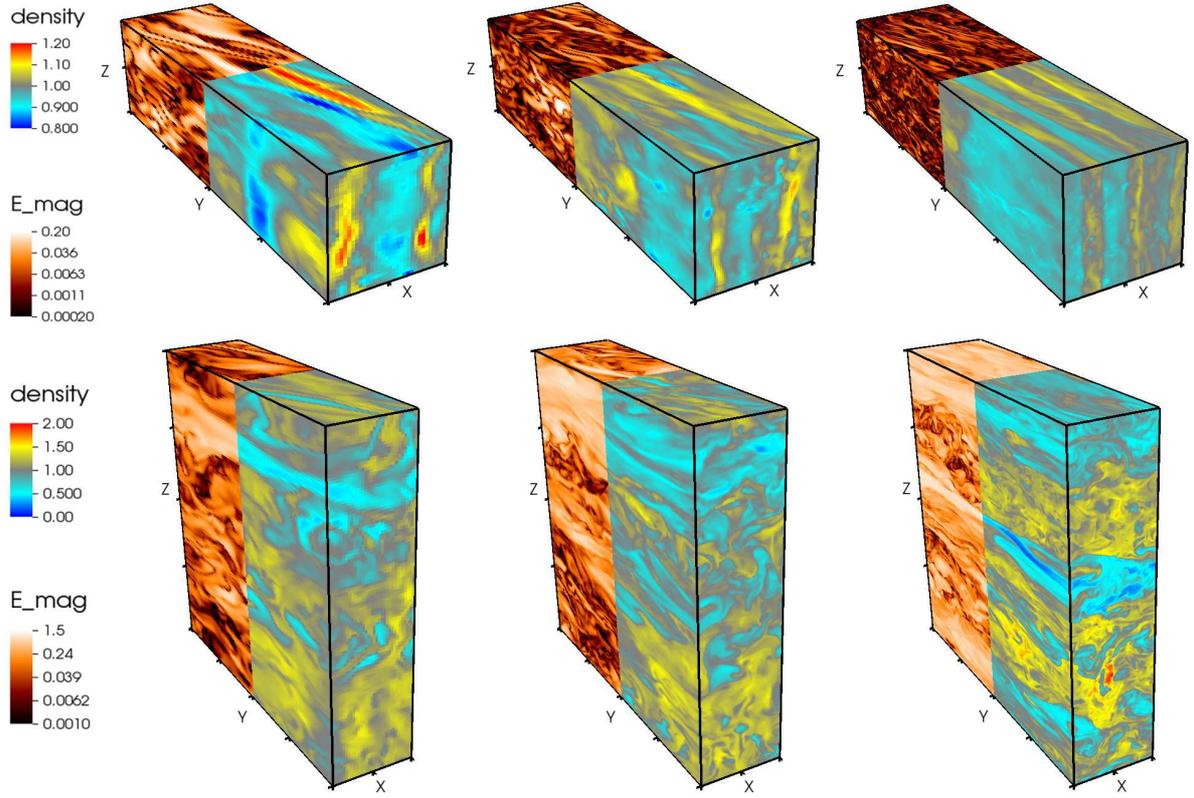}
\caption{\small{Snapshots (at $t=150$ orbits) from standard box (top row,
$\lx=\ly/4=\lz=1\,H$) and tall box (bottom row, $\lx=\ly/4=\lz/4=1\,H$) simulations.
Shown in each case are
magnetic energy (far half) and density (near half) distributions. From left to
right, the resolution is $32/H$, $64/H$, and $128/H$.  Turbulence becomes weaker in
the standard
box case as resolution increases; while it maintains constant amplitude in the tall
box runs.  Note the color scale of the magnetic energy is logarithmic,
and the color bars in the bottom and top rows are different.  }}
\label{fig:dm}
\end{figure*}

\subsection{Saturation in tall versus short boxes}
\label{sec:convergence}
\begin{figure*}
	\includegraphics[width=8cm]{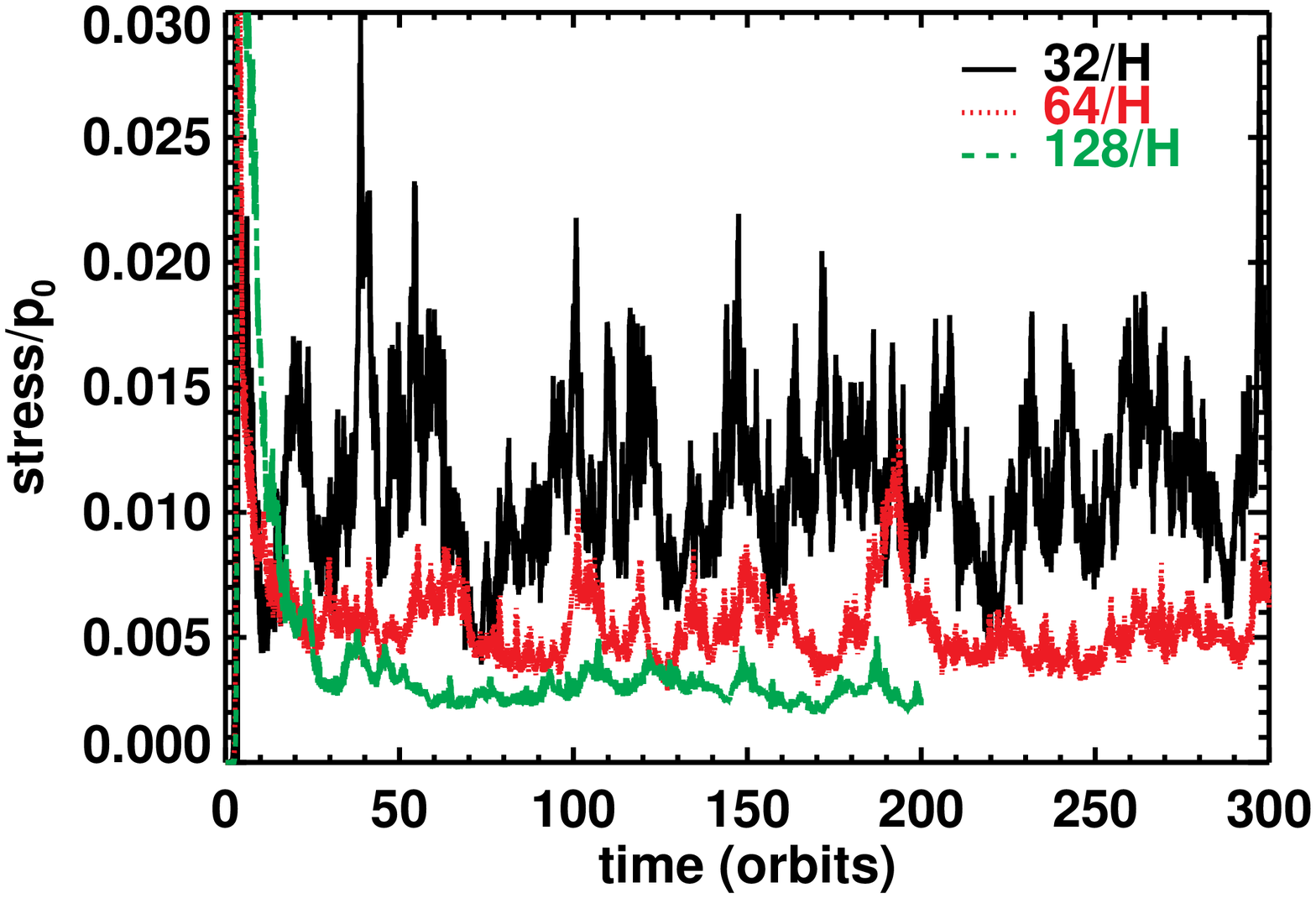}
	\includegraphics[width=8cm]{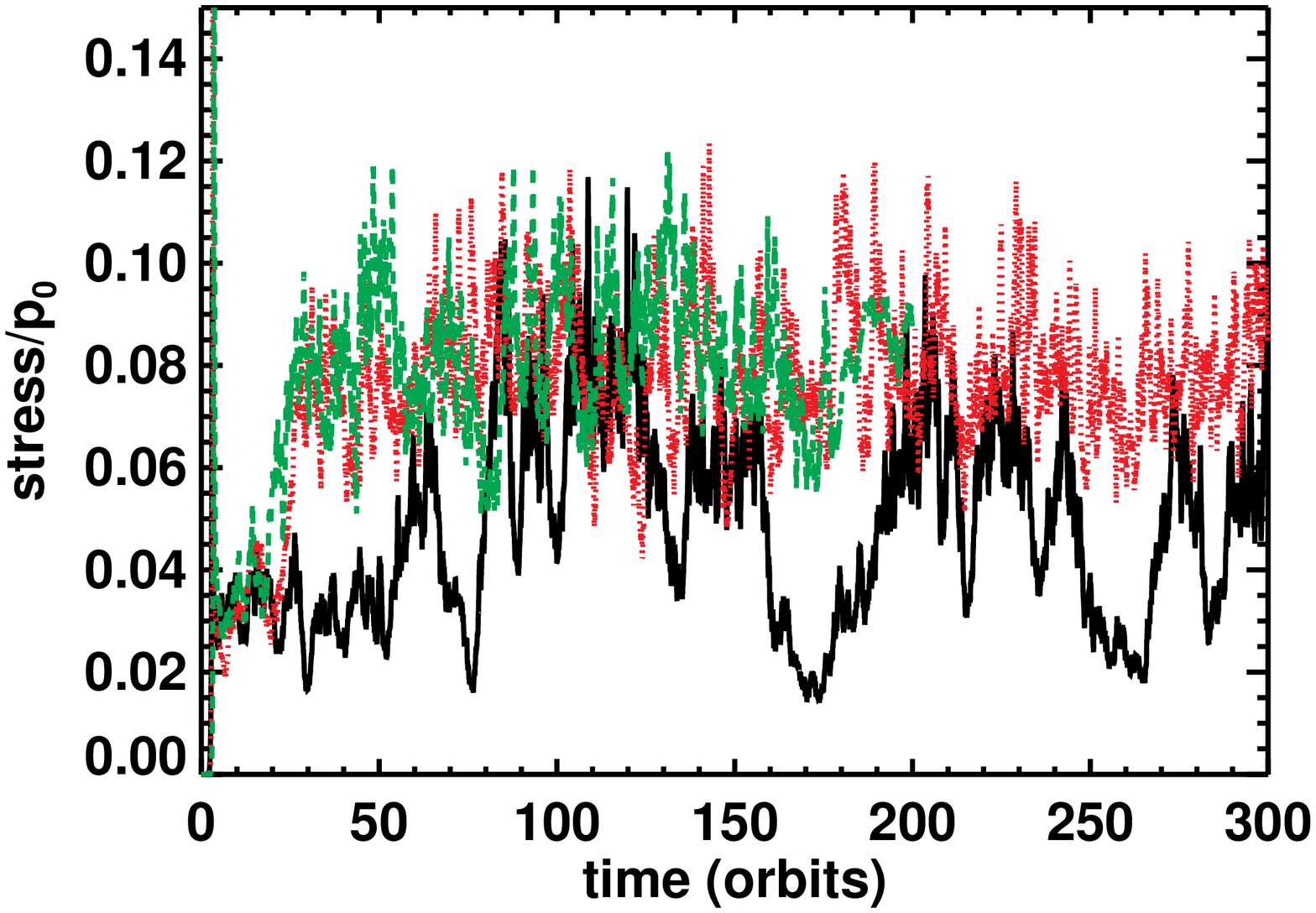}
\caption{\small{The volume averaged stresses
		for a standard box ($(\lx,\ly,\lz) = (1,4,1)\,H$, left panel)
		and a tall box ($(\lx,\ly,\lz) = (1,4,4)\, H$, right) with various
		resolutions: $32/H$(black solid),$64/H$(red dotted) and $128/H$(green dashed). 
		In contrast to the small box, the vertically elongated box achieves good convergence. }}
\label{fig:convergence}
\end{figure*}

We begin by studying saturation in the standard computational domain,
$(\lx,\ly,\lz) = (1,4,1)\,H$.
We begin with a numerical resolution of $32/H$.  We refer to this simulation as x1y4z1r32,
where our naming convention gives both the box dimensions (in $H$) and resolution.  We then
double this resolution twice
(simulations x1y4z1r64 and x1y4z1r128) to reproduce previous results.
The top panels of Figure~\ref{fig:dm} show 3D snapshots of the magnetic energy
and density in these runs.   As the resolution improves from left to right, the
magnetic energy is reduced, and turbulence becomes weaker.  At the highest resolution
(x1y4z1r128 in the right panel) the density field is dominated by $k_{\rm z}= 0$
shearing waves \citep{HP2009}, and there is little indication of turbulence. 

We then repeat these runs in a taller box, $(\lx,\ly,\lz) = (1,4,4)\,H$, varying the
resolution from $32/H$ for run x1y4z4r32, to $64/H$ for run x1y4z4r64,
and to $128/H$ for run x1y4z4r128.
Snapshots of magnetic energy and density for these runs are shown in the bottom row of
Figure~\ref{fig:dm}. In contrast to the standard box case, we see no signs of weakening turbulence as
resolution improves from left to right. The typical magnetic energy rises slightly from 
x1y4z4r32 to x1y4z4r64, and stays roughly at the same level as the resolution doubles again in
x1y4z4r128. The typical density fluctuation follows the similar trend, and are dominated by
small scale turbulent fluctuations.  At the same time, both the density and magnetic field
show a large amplitude vertical mode with wavelength about half the vertical size of the box. 

To quantify the resolution effects, we measure the volume averaged total stress $\alpha_{\rm tot}$
(the sum of the Maxwell and Reynolds stresses normalized with thermal pressure defined in
section~\ref{sec:diag}) and the results are shown in
Figure~\ref{fig:convergence}.  In the standard box runs (the top left panel), after a short transient
growth ($\sim 30$ orbits), the total stresses reaches a quasi-steady state that last hundreds of orbits
through the end of the simulations.   The 
time-averaged stresses are therefore well defined
and are measured over the last $100$-$200$ orbits (see Table-\ref{tab:tab1}
for the exact number used for averages). Comparing the saturation levels at different
resolutions, 
we confirm the stress decreases as the resolution improves, roughly by a factor of two
every time the resolution is doubled.  In particular, $\alpha_{\rm tot}$ drops from
$10^{-3}$ for x1y4z1r32 to $5\times 10^{-4}$ when the resolution is doubled, and drops by another factor of
$\sim 2$ when $128/H$ is used.  This behavior is identical to that previously reported
\citep[e.g.,][]{FP2007,simonetal2009,Guan2009,bodoetal2011}.

Remarkably, the decrease in stress with resolution that appears in the standard box is
not reproduced in
the taller box simulations. In the bottom left panel of
Figure~\ref{fig:convergence}, we show that the volume averaged stress does in fact converge
with numerical resolution for runs x1y4z4r32 through x1y4z4r128.
In these cases, $\alpha_{\rm tot}\simeq 0.08$ is reached once the resolution exceeds $32/H$. 

Figure ~\ref{fig:convergence} also demonstrates that the amplitude of the converged
stress in the tall box is
significantly greater than those in the standard domain.
For example, the time averaged $\alpha_{\rm tot}\simeq 0.079$ in x1y4z4r64, almost $15$ times
larger than $\alpha_{\rm tot}\simeq 0.005$ in x1y4z1r64, while $\alpha_{\rm tot}\simeq 0.08$ in x1y4z4r128, some
$\sim 30$ times greater than that of x1y4z1r128.
The magnetic energy in simulations performed in the tall box is also
much greater than in the standard box runs; typically $\langle B^2/8\pi P_0\rangle \simeq 0.27$ which is a factor of
$\gtrsim 50$ time bigger (this is also evident in Figure~\ref{fig:dm} where different color
scales must be used for the two cases). 
The large values of the stress achieved in taller boxes is also of interest
in that they are
close to the values inferred from observations suggesting that in fully ionized
accretion disk (e.g. dwarf novae in outburst) $\alpha_{\rm tot}\sim 0.1$--$0.4$, alleviating
some of the concerns raised by
\citet[][]{kpl2007}
regarding the discrepancy between values suggested by observations
and measured in simulations.

\begin{figure*}
	\includegraphics[width=8cm]{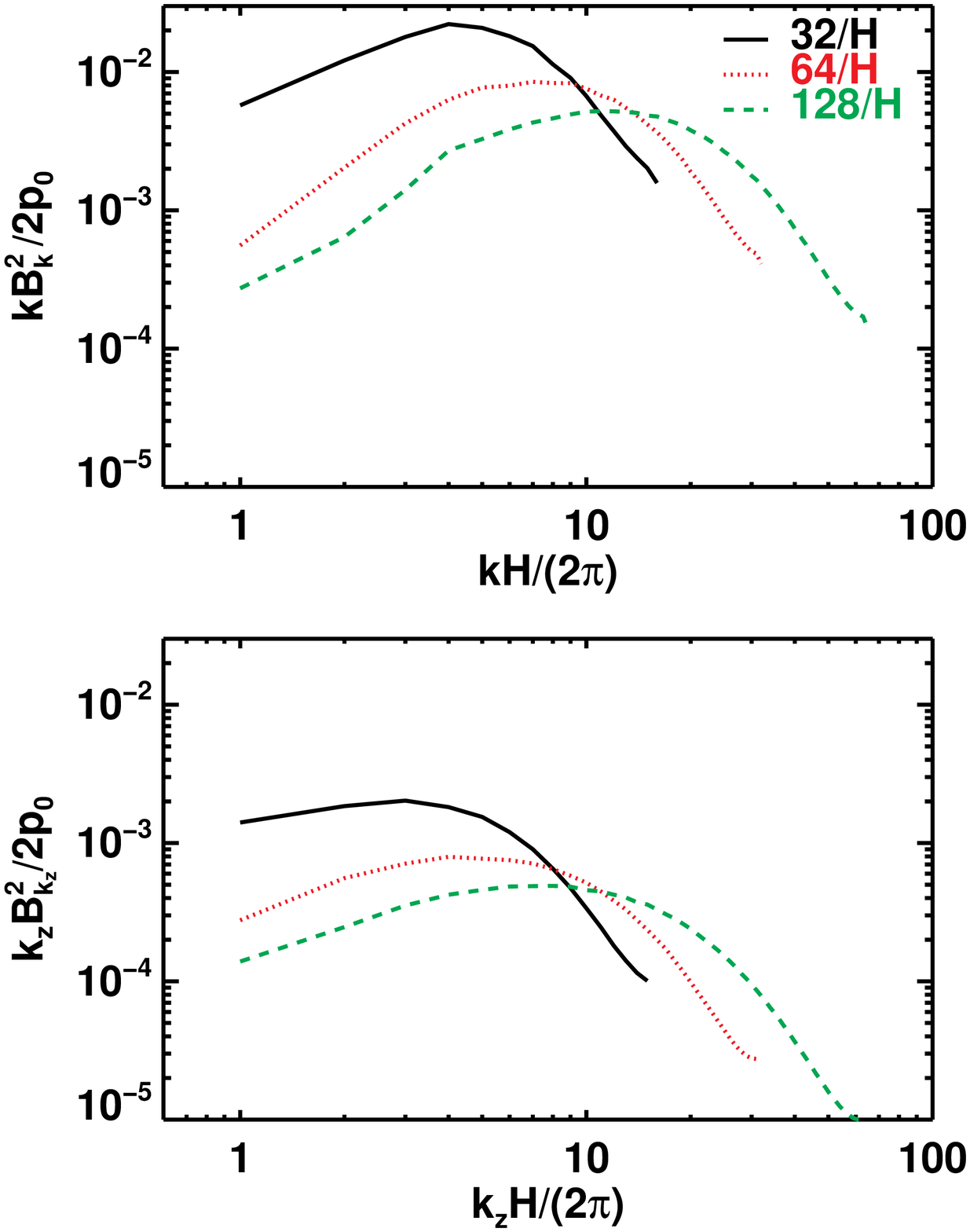}
	\includegraphics[width=8cm]{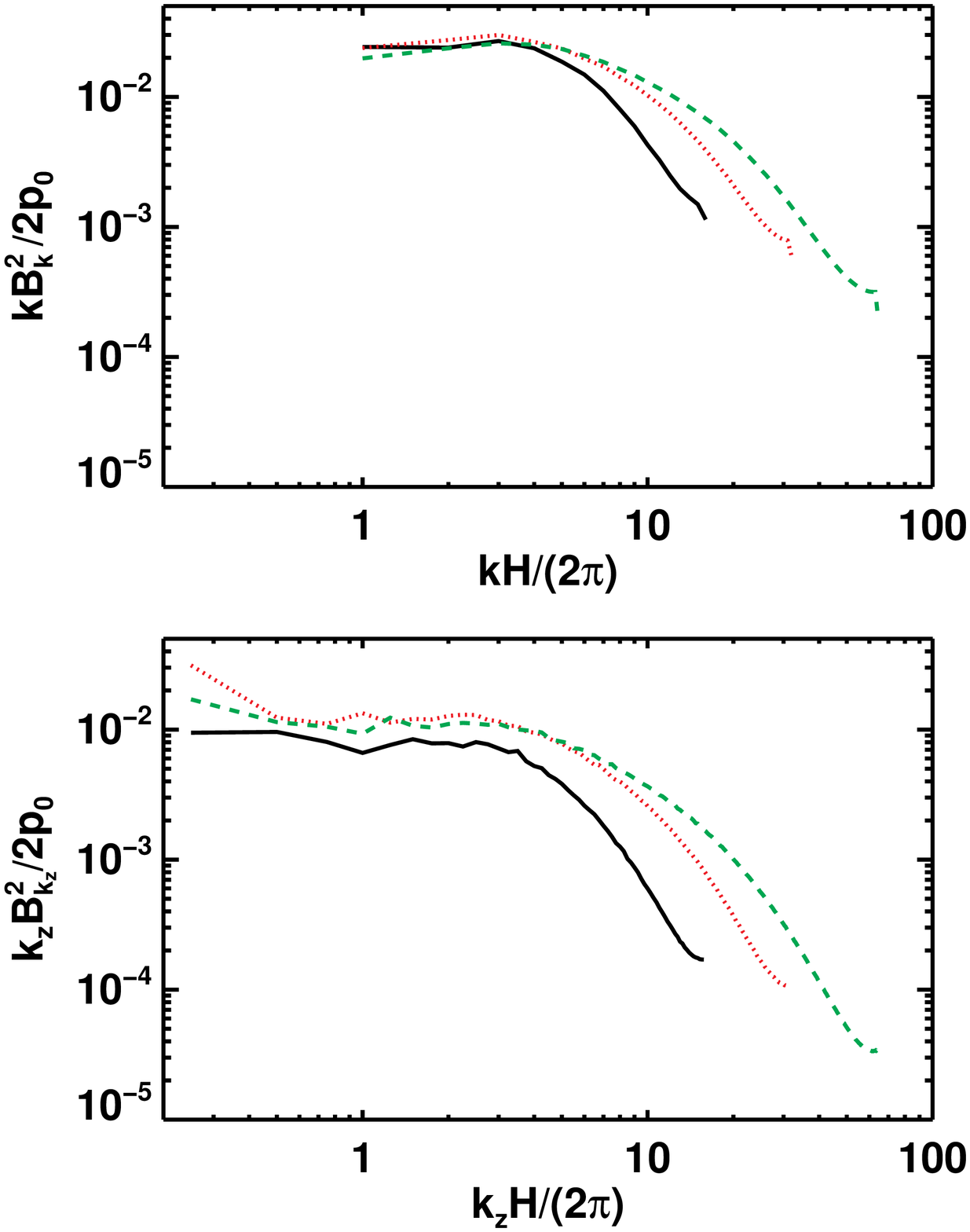}
\caption{\small{The power spectra of magnetic density
		energy for a standard box ($(\lx,\ly,\lz) = (1,4,1)\,H$, (left two panels)
		and a tall box ($(\lx,\ly,\lz) = (1,4,4)\, H$, right two panels) with various
		resolutions: $32/H$(black solid),$64/H$(red dotted) and $128/H$(green dashed), where
		$B_k^2$ is spherical shell integrated value and $B_{k_z}^2$ is integrated along
		constant $k_z$ plane. In contrast to the small box case, where the self-similar
power spectra decays with resolution, it achieves good convergence at low wave number in the tall
box. }}
\label{fig:convergence_pwr}
\end{figure*}

\begin{figure*}
	\includegraphics[width=16cm]{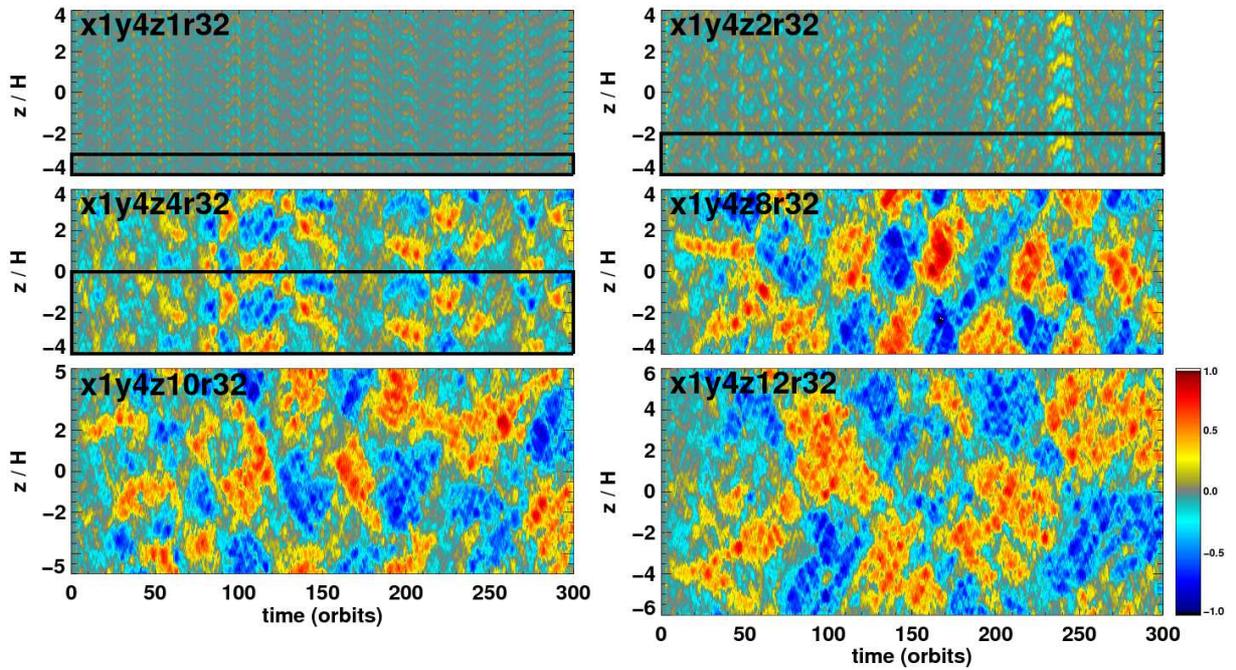}
\caption{\small{
Space-time diagrams of the horizontally averaged azimuthal field
$\ob{B}_y$ for various
runs as labeled. Runs with $\lz < 8\,H$ are duplicated $8$ (for
x1y4z1r32), $4$ (for x1y4z2r32), and $2$ (for x1y4z4r32) times, so that the images have
the same aspect ratio of run x1y4z8r32. The individual black box
shows the actual box we simulate. Large and structured
azimuthal B-field appears when $\lz >2\,H$.} }
\label{fig:zt_cycle}
\end{figure*}
\begin{figure*}
	\includegraphics[width=16cm]{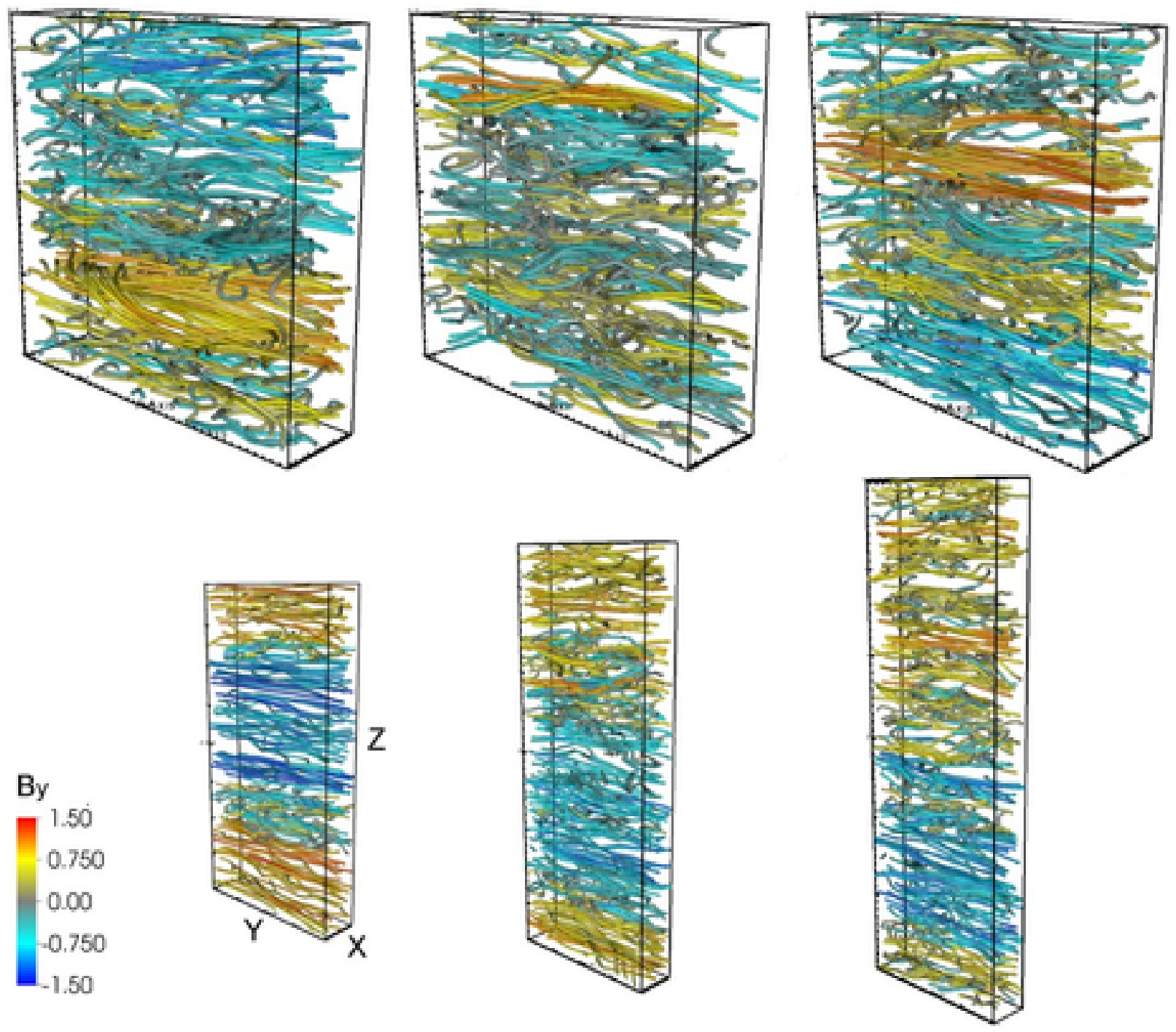}
\caption{\small{Top row: large scale magnetic field in x1y4z4r32 run reverses between $t=200$ orbits
(top left) and $220$ orbits (top right); $t=210$ at the middle show more tangled field during this
transition. Bottom row: snapshots of large scale field structures in other tall box runs (x1y4z8r32, 
x1y4z10r32, and x1y4z12r32 from left to right). In all panels, tubes are field lines in the box with 
starting points randomly distributed; color bar describes the value of $B_y/\sqrt{4\pi\rho_0}\cs$ to show the orientation 
and strength of magnetic field in azimuth.}}
\label{fig:field_line}
\end{figure*}

Further important insights into the differences between the saturated state in standard
and tall boxes are revealed by the Fourier power spectrum.
The right column of 
Figure~\ref{fig:convergence_pwr} shows the magnetic energy density power spectra calculated over either 
spherical shells of constant $k$ ($k B_k^2$) or along a constant $k_z$ plane ($k_z B_{k_z}^2$) for
both the standard (left) and tall box (right). The definition of both can be find in
\citet[][section~2.1]{davisetal2010}. The spectra in the standard box peak 
at an intermediate wavenumber that increases with the resolution, and with an overall
amplitude that decreases with resolution (but with a shape that is unchanged).
In stark contrast, however,
in the tall box the power spectra peak (or plateau) at the lowest wavenumbers
($k \sim 1$-$5$) at all resolutions.  Such spectra are much more reminiscent of
turbulence driven at a large outer scale, with an energy cascade to high $k$ independent 
of numerical resolution.  These spectra suggest there is a resolved characteristic outer 
length scale in the tall box simulations.
We note the spectrum and its dependence on resolutions of those tall box runs are also very
	different than those using standard short boxes with explicit dissipation \citep[e.g.,
	Figure~3 in][]{Fromang2010} due to the lack of a large scale dynamo in the short boxes.
	However, a large scale dynamo, as we will show in the next subsection, is present in our
tall box simulations.

\subsection{A large scale dynamo}
\label{sec:dynamo_existence}
The images of magnetic energy and density shown in Figure~\ref{fig:dm},
and the power spectra in Figure~\ref{fig:convergence_pwr},
both indicate that large-amplitude and large-scale vertical structure
appears in simulations in tall boxes.  In this section, we show this
structure is associated with a {large scale} dynamo action that is triggered in the
vertically extended box.
Figure~\ref{fig:zt_cycle}
plots a space-time diagram of the horizontally averaged azimuthal
magnetic field in both standard and tall boxes.  Patches of strong
$\ob{B}_y$ that extend over vertical regions of size $\sim H$ are
evident in run x1y4z4r32, with the sign of the field alternating
quasi-periodically over $\sim 10$-$20$ orbits.  The pattern become
even more clear in the panels corresponding to runs x1y4z8r32, x1y4z10r32 and
x1y4z12r32 (see especially the middle right panel in
Figure~\ref{fig:zt_cycle}).  Clearly, a large scale dynamo must be operating
that generates such strong ordered toroidal field.  The 3D structure
of the magnetic field at three different times during one reversal
cycle in run x1y4z4r32 is explored further in Figure~\ref{fig:field_line}.
Large scale ($k_x = k_y = 0$ and $k_z=1$) patterns in the azimuthal
magnetic field form along the azimuthal direction.  Spatially, the
field lines flip signs (from red to blue) with height to conserve
the zero-net-flux initial condition. Their orientations also reverse
from one state of coherent structures (top left at $t=200$ orbits)
to another (top right at $t=220$ orbits).  This result is similar
to the dynamo cycle reported in \citet[][]{LO08} (see their Figure~3).
However, we emphasize that no explicit dissipation (viscosity or
resistivity) is required to capture the dynamo, nor is it required
to see a converged level of stress.

The large scale dynamo maintains a time averaged mean field 
$\langle\langle \ob{B}^2/8\pi P_0\rangle\rangle_t\simeq 0.06$ for x1y4z4r32 and $\simeq 0.13$ for
x1y4z4r64 and x1y4z4r128 runs, which amount to $\sim 40$-$50\%$ of the total magnetic energy in the
box (see Table~\ref{tab:tab1} and ~\ref{tab:tab2}), close to a state of equalized mean
$\ob{\mathbf{B}}$ and turbulent ($\mathbf{b}$) field strength. 
In contrast, runs in the standard box do not exhibit strong cyclic dynamo behavior (see top left
panel of Figure~\ref{fig:zt_cycle}), and the mean field is rather weak,
$\langle\langle \ob{B}^2/8\pi P_0\rangle\rangle_t \lesssim 0.1 \langle\langle b^2/8\pi
P_0\rangle\rangle_t$ for x1y4z1r32, and it drops further down to $\simeq 2\%$ of the turbulent
magnetic energy in run x1y4z1r128.

Why does dynamo action produce a larger value for the stress which does not vary
with numerical resolution?
The large scale vertical patches of azimuthal magnetic field produced by the dynamo act 
locally as a region with net toroidal flux.  Thus, each $\lz \sim H$ patch acts as
an unstratified shearing box {\em with net toroidal field}.  As is already known, 
shearing boxes with net flux produce saturated stress which is independent of
resolution \citep{Guan2009,SH2009}.
As a result of (locally) strong magnetic field, the saturated Maxwell stress in x1y4z4r128 is $\am\simeq
0.0714$, $\sim 31 \times$ greater than that of x1y4z1r128. Of this total, $\sim 16\%$ is due to the
correlated mean field $-\ob{B}_x\ob{B}_y/4\pi$, while the majority is still from the
correlation
between the perturbed field components $-b_x b_y/4\pi$ (since $\ob{B}_x$ is still small).
In contrast, in the standard-sized box almost all of the stress 
is associated with the perturbed field.

We find the stress-to-energy ratio, $\alpha_{\rm mag}
\equiv \langle \langle -2 B_x B_y \rangle\rangle_t/ \langle\langle B^2\rangle\rangle_t \simeq 0.27 $ in the tall
box, smaller than in the standard box case in which $\alpha_{\rm mag} \simeq 0.46$. A diminished $\alpha_{\rm mag}$
is also observed in simulations with strong net azimuthal flux \citep[run Y8 in][]{hgb95}, where the imposed
azimuthal mean field resembles a subsection of our box which contains $\ob{B}_y$ of the same sign.

\subsection{Varying the aspect ratio $\lz/\lx$: a parameter survey}
\label{sec:aspect}
\begin{figure}
	\includegraphics[width=\columnwidth]{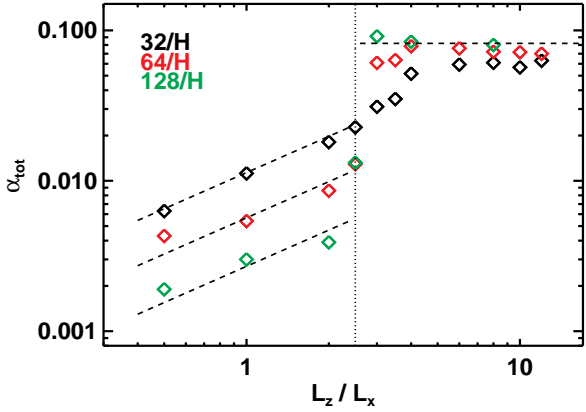}
\caption{\small{The sum of Maxwell and Reynolds stress for various aspect ratios ($\lz/\lx$) and
		resolutions for $\lx = \ly/4 = 1\, H$ boxes. The vertical dotted line separates the `diverging' and
`converging' regions: to the left of this line, the stress decays with increasing resolution; the stress
converges to $\sim 0.08$ for different $\lz/\lx$ on the right. }}
\label{fig:alpha_aspect}
\end{figure}

We now investigate the effect of varying the aspect ratio using
$\lz/\lx \in \{0.5,1,2,2.5,3,3.5,4,6,8,10,12\}$ with fixed
size $\ly=4H$ and $\lx=H$ in the horizontal dimensions.
For each of $\lz/\lx$, we also vary the resolution to study numerical
convergence. We first plot the $\alpha_{\rm tot}$ in Figure~\ref{fig:alpha_aspect}.  This
figure clearly demonstrates what may be our most important result.  The data
falls into two groups which show distinctly different behavior; the two groups are
separated by the vertical dashed line at $\lz/\lx \simeq 2.5$. For
aspect ratios $\lesssim 2.5$ (to the left of the dotted line), the stress decays linearly
with the vertical size $\lz$.  Moreover, at any given aspect ratio, the stress decreases with
increasing numerical resolution as shown by the decreasing amplitude of the black ($32/H$),
red ($64/H$) and green ($128/H$) points.  Clearly the standard box, with $\lz/\lx=1$ falls in
this group.
In contrast, for aspect ratios $\lz/\lx \gtrsim 2.5$ (to the
right of the dotted line), the saturated stress associated with MRI turbulence appears
independent of the vertical size $\lz$, 
approaching $\alpha_{\rm tot} \sim 0.1$.  Moreover, numerical convergence in the value of the stress is achieved
for aspect ratios $\lz/\lx>4$.  For example, for $\lz/\lx=8$, the stress values are nearly
identical for resolutions of $32/H$ through $128/H$.
The behavior of the two groups in this figure simply reflect the emergence of dynamo action
in runs with a large vertical extent, which controls the
stress.   For example, the $\lz/\lx = 4$, $8$, $10$, and $12$ cases in
Figure~\ref{fig:zt_cycle} and \ref{fig:field_line} show strong cyclic azimuthal magnetic field
driven by the underlying dynamo mechanism.  

\begin{figure}
	\includegraphics[width=\columnwidth]{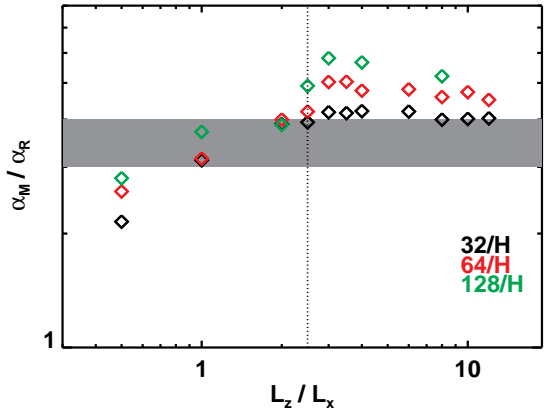}
\caption{\small{The ratio of the Maxwell stress and the Reynolds stress as a function of $\lz/\lx$
	and numerical resolution. The vertical dotted line at $\lz/\lx = 2.5$ same as
Figure~\ref{fig:alpha_aspect}. The Maxwell stress always dominates the Reynolds in all runs; the
ratio increases by a factor of $\sim 2$ in the small box regime, while stays roughly constant in the
tall box cases. The tall box runs have greater $\alpha_{\rm M}/\alpha_{\rm R}$ than previous
measurements marked as the shaded region. }}
\label{fig:m2r}
\end{figure}

For all runs, the Maxwell stress dominates the total turbulent angular momentum transport; the
ratio of the Maxwell to the Reynolds stress varies with different $\lz/\lx$ in a similar way
(but at reduced amplitude) 
as the $\alpha_{\rm tot}$. In Figure~\ref{fig:m2r}, we find $\am/\ar$ rises gradually from $\sim 3$
to $4$ as $\lz/\lx$ is increased from $0.5$ to $2$. The ratio then levels off, ranging between $4$ and
$6$. This is greater than the typical values previously reported in the literature with Keplerian
shear ($\sim 3$-$4$) \citep[][]{hgb95,abl1996,shgb96,hbw1999,sano2004,pessahetal2006}. As the strong coherent
field structures tend to eliminate strong velocity fluctuations and therefore reduce the Reynolds
stress in those regions (similar to those observed in the corona regions of stratified boxes
\citep[][]{Brandenburgetal1995,MS2000}), the box averaged $\am/\ar$ therefore rises above the
previously reported values.

\begin{figure}
	\includegraphics[width=\columnwidth]{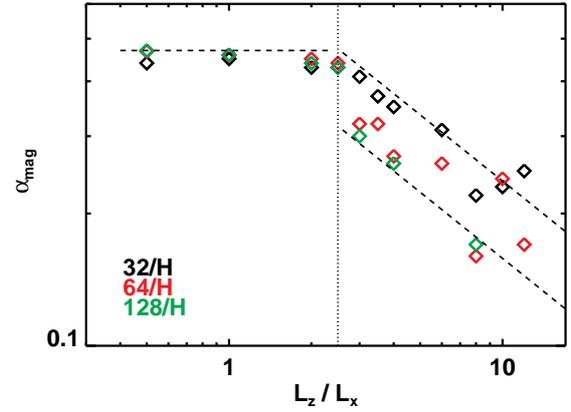}\\
	\includegraphics[width=\columnwidth]{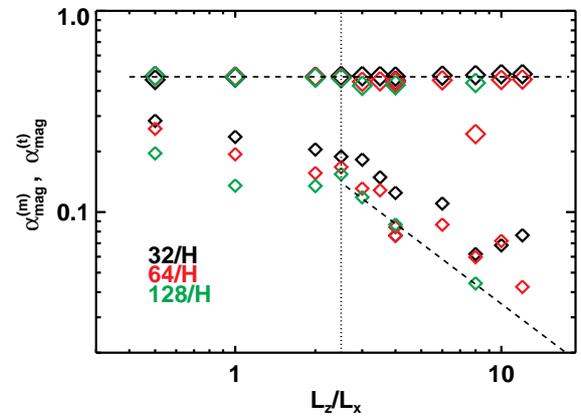}
\caption{\small{Top: $\alpha_{\rm mag}\equiv -2\langle\langle B_x B_y\rangle\rangle_t / \langle\langle
B^2 \rangle\rangle_t$ for various $\lz/\lx$ and resolutions. Again, the vertical dotted line
delineates the short and tall box regimes. The horizontal dashed line shows that $\alpha_{\rm mag}$
stays constant in the small box regime; the dashed lines at $\lz/\lx > 2.5$ follows the $(\lz/\lx)^{-1/2}$ 
power law. Bottom: stress-to-energy ratios computed with pure mean field (small symbols) and
turbulent field (large symbols) as defined in \ref{eq:alpha_mag_bn}. The turbulent ratio keeps
constant $\alpha_{\rm mag}^{(t)}\sim 0.47$ (the horizontal dashed line); while its mean field counterpart $\alpha_{\rm
mag}^{(m)}\propto (\lz/\lx)^{-1}$ (the dashed line at $\lz /\lx >2.5$), much smaller than $\alpha_{\rm mag}^{(t)}$. }}
\label{fig:amag}
\end{figure}

The ratio between the Maxwell stress and magnetic energy, $\alpha_{\rm mag}$ (see definition
in section~\ref{sec:dynamo_existence}), is usually adopted as a measure of sufficient 
resolution to capture MRI turbulence \citep[e.g., $\alpha_{\rm mag}=0.3-0.4$ in][]{blackman2008,hgk2011}.  
We find consistent results for our standard boxes ($\lz/\lx \lesssim 2.5$), in which a relatively constant 
$\alpha_{\rm mag} \sim 0.47$ is obtained, in spite of that no convergence is achieved for the
stress.  Surprisingly, this ratio falls with $(\lz/\lx)^{-1/2}$ in our tall box runs where
converged stresses are found. As the stress stays roughly constant in those tall boxes,
it is the increase of magnetic energy (or $1/\beta$) as $(\lz/\lx)^{1/2}$ that drives this
scaling.

The dynamo effect in tall boxes alters the underlying magnetic field structure.
As discussed 
in section~\ref{sec:convergence}, most of the stress comes from the correlation of the small scale field, 
and $\langle -\Bx \By \rangle \sim  \langle -\bx \by\rangle \gg \langle -\ob{B}_{\rm x}\ob{B}_{\rm
y}\rangle$ holds true for both groups.
However, the magnetic energy in smaller boxes is mostly from the azimuthal component of the 
turbulent field, i.e., $\langle B^2\rangle \sim \langle \by^2 \rangle \gg \langle \ob{B}_{\rm y}^2\rangle$; 
while $\langle B^2\rangle \sim \langle \ob{B}_{\rm y}^2\rangle \gg \langle \by^2\rangle$ in tall boxes. 
As a result, $\alpha_{\rm mag}$ roughly measures $\left|\bx/\by\right|$ in the standard box cases, but it traces 
a very different quantity, $\left|\bx \by\right |/\ob{B}_{\rm y}^2$ in the tall boxes. The effects of 
increasing the aspect ratio in an already elongated box would only introduce a stronger mean
magnetic field. For instance, we find the total magnetic energy rises from $\langle B^2/8\pi
P_0\rangle \simeq 0.27$ to $0.37$ when comparing x1y4z4r128 and x1y4z8r128 in Table~\ref{tab:tab2},
and the energy increase mostly goes into $\langle\ob{B}_{\rm y}^2/8\pi P_0\rangle$ so that the latter increases 
from $\simeq 0.13$ to $0.25$ while the other magnetic field components stay constant.

In a recent review, \citet{BN2015}
pointed out the stress-to-energy ratios for the pure mean and turbulent field could be different.
Following their proposal, we define
\beq
\alpha_{\rm mag}^{(m)}\equiv -2\frac{\langle\ob{B}_{\rm x}\ob{B}_{\rm y}\rangle}{\langle
\ob{B}^2\rangle} \,; 
\alpha_{\rm mag}^{(t)}\equiv \frac{-2 \langle \bx \by \rangle}{\langle B^2\rangle -\langle
\ob{B}^2\rangle}
\label{eq:alpha_mag_bn}
\enq
for the mean field and turbulent field respectively. In the bottom panel of Figure~\ref{fig:amag},
we find the ratio of the turbulent field (larger symbols) still reaches a constant $\alpha_{\rm
mag}^{(t)}\sim 0.47$, while the mean field counterpart $\alpha_{\rm mag}^{(m)}\ll \alpha_{\rm
mag}^{(t)}$, and decreases linearly with $\lz/\lx$. These are consistent with the results found in
\citet{BN2015}.

\begin{figure*}
	\includegraphics[width=8cm]{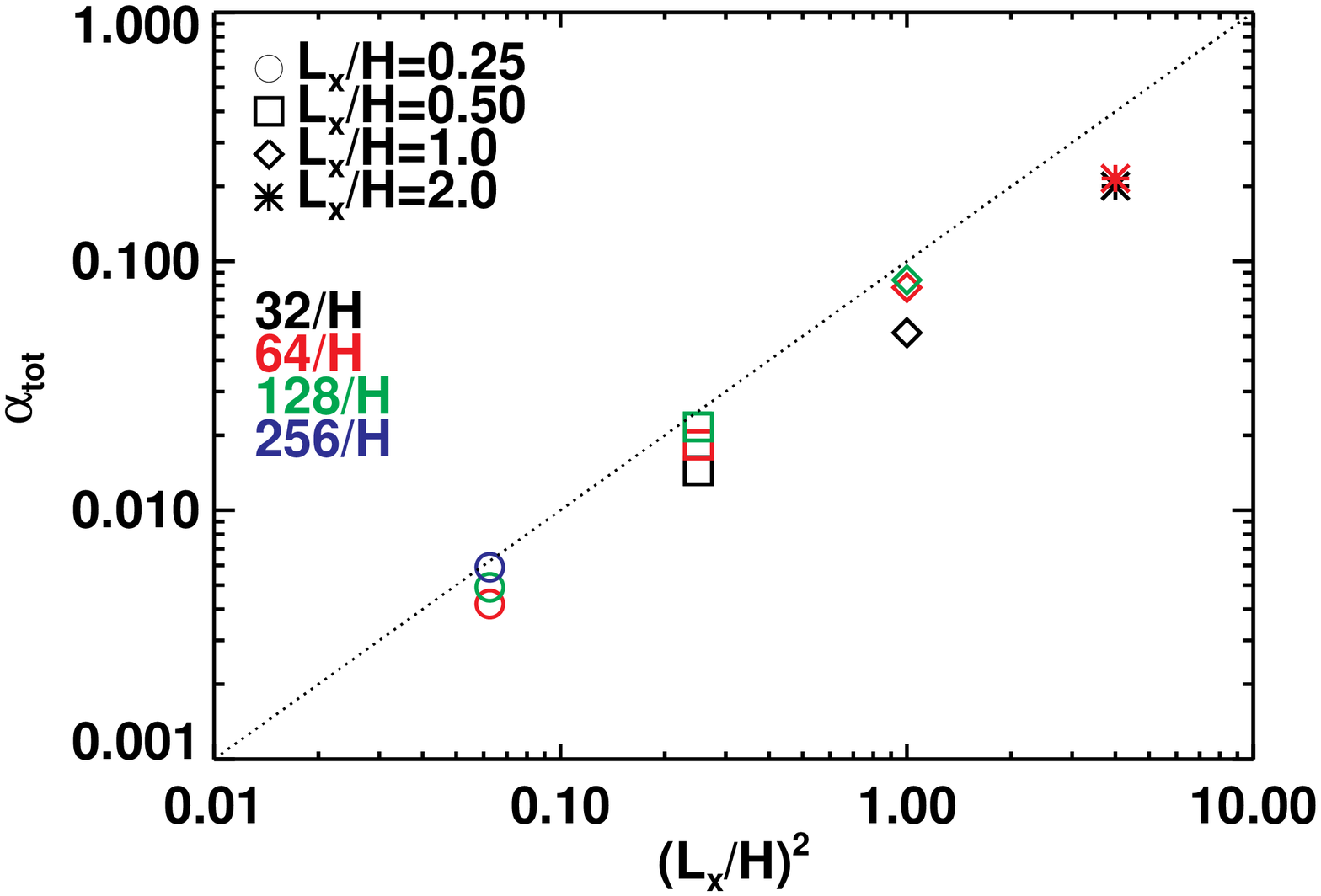}
	\includegraphics[width=8cm]{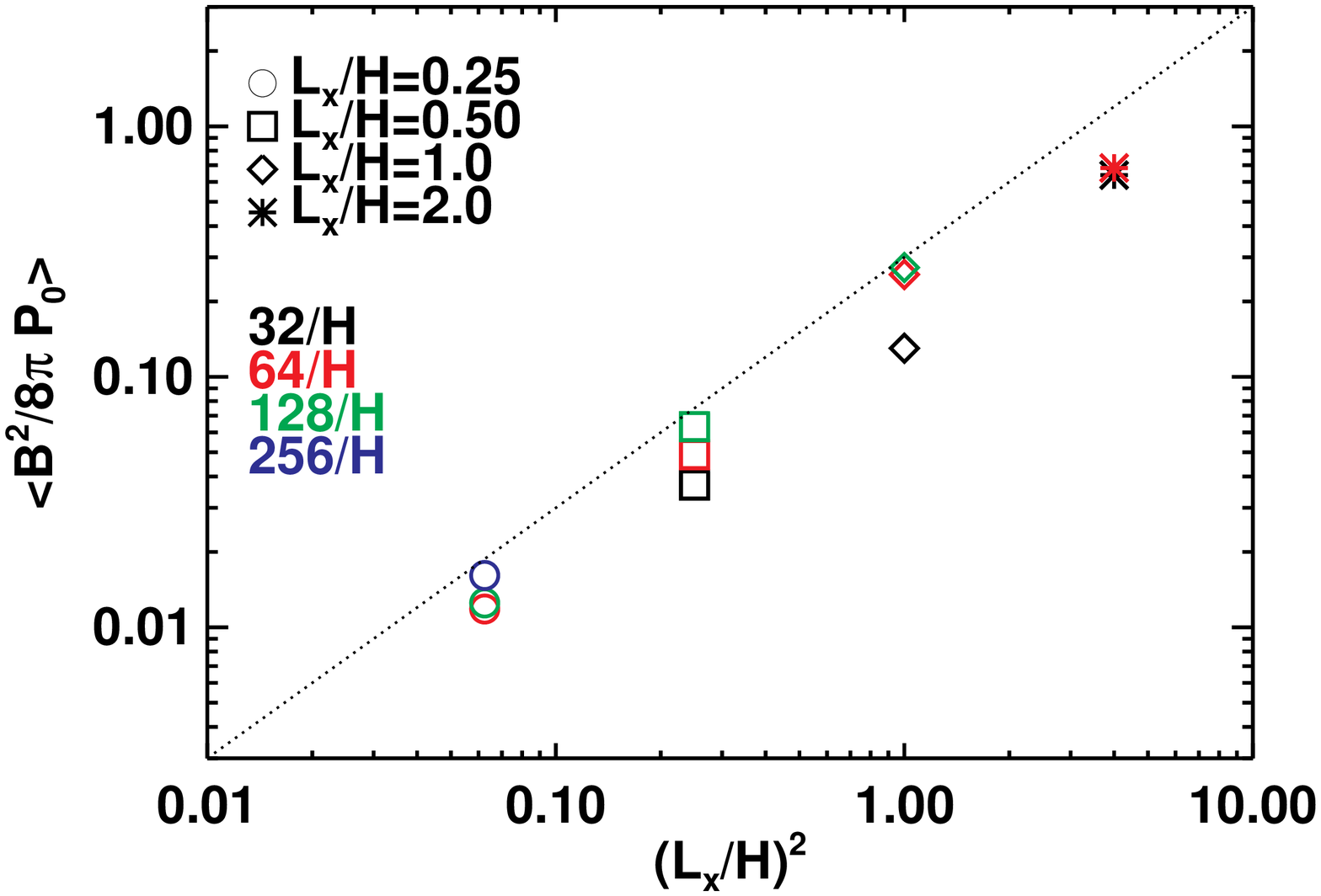}
\caption{\small{The box size dependence of the stress (left) and magnetic energy (right) for $L_z/L_x=4$ runs. 
We note that for both panels, the $x$-axes are $(\lx/H)^2$ instead of $\lx$. The dashed lines shows
the $\propto (L_x/H)^2$ scaling. A strong linear correlation between the stress/energy and
$(\lx/H)^2$ is obtained which indicates a way to re-normalize $\alpha$ and $\langle
B^2\rangle$ using the squared length scale $\lx^2$.}}
\label{fig:alpha_aspect4}
\end{figure*}

\subsection{Varying the box size with fixed aspect ratio}
\label{sec:size}
For given box size $\ly=4H$, $\lx=H$ in the horizontal domain,
we find the stresses converge in the tall box runs ($\lz/\lx \gtrsim 2.5$)
owing to the emergence of large scale azimuthal magnetic field produced by dynamo action
that acts as a local non-zero net flux. 
Since the mean field sets an extra length scale that is directly related to the box size, it is also
important to see how does the stress depend on the box size itself.  Therefore we have
performed an additional series of runs in which we vary $\lx$, but keep the
aspect ratio $\lz/\lx$ fixed. 

In general, we find the amplitude of the saturated stress still converges to a constant value
independent of numerical resolution as long as $\lz/\lx > 2.5$
(see Table-\ref{tab:tab1}).
However, the value to which the stress converges depends on the box size, as found previously
for shearing box simulations with net azimuthal field \citep{hgb95,Guan2009}.  As shown in 
Figure~\ref{fig:alpha_aspect4}, both $\alpha_{\rm tot}$ and $\langle B^2/8\pi P_0\rangle$, measured from runs 
with aspect ratio of $1:4:4$ scale as $\propto \lx^2$.  This differs from the saturation predictor used in
\citet{hgb95}, in which a linear relation $\propto \ly$ is reported.  However, their
relation applied to
relatively weak ($\beta\sim O(10^2)$) external azimuthal field is very different
than the very strong ($\beta\sim O(10)$), dynamo generated oscillating azimuthal field observed in our runs.

In fact, the unstratified box does not have any intrinsic length scales other than the box
size $\lx$, cell width $\Delta x$, and (for compressible flows) the sonic scale $c_s/\Omega$.
If we assume that compressible effects can be ignored (we return to this point below),
then based on a dimensional analysis the stress can be normalized by $\rho_0(q\Omega\lx)^2$
\citep{FP2007,Guan2009}, giving stress $\propto \lx^2$ as reported above.  As long as the
simulation is resolved $\lx /\Delta x \gtrsim 32$, we find negligible 
dependence of the stress on
numerical resolution, e.g. Figure~\ref{fig:alpha_aspect4}. 
We find similar scaling with size for other quantities such as magnetic and
kinetic energy, and Reynolds stress.  In addition, the scaling $\propto \lx^2$ applies to
runs using other values of the 
aspect ratio, e.g., $\lz/\lx=8$ as tabulated in Table~\ref{tab:tab1}.

\begin{figure}
	\includegraphics[width=\columnwidth]{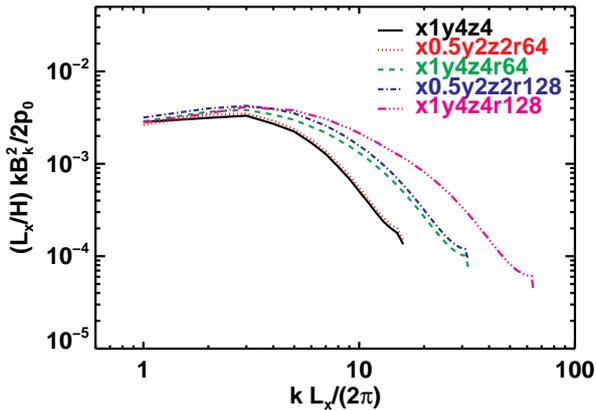}
\caption{\small{Rescaled magnetic energy power spectra for boxes with same aspect ratio $\lz/\lx=4$.
The close matches between x1y4z4 and x0.5y2z2r64, x1y4z4r64 and x0.5y2z2r128 indicate that 
the underling turbulence is intrinsically the same but only rescaled to a different box size.}}
\label{fig:pwr}
\end{figure}

In Figure~\ref{fig:pwr}, we plot the Fourier power spectra of five different
runs computed with different numerical resolutions and physical box size (both $\lx=H$ and
$\lx=0.5H$), but normalizing both the spectra and wavenumber according to the above scaling.
Interestingly, all runs contain the same amplitude and slope at small
$k$ (the ``inertial range"),
but extend to successively larger $k$ as the resolution is increased.  Moreover, 
run x0.5y2z2r64 recovers the results of run x1y4z4r32 identically, and run
x0.5y2z2r128 resembles x1y4z4r64 as well.  
Since these two pairs of runs have the same range of scales $\lx/\Delta x$, we expect them
to show similar behavior provided there are no intrinsic length scales in the model.
The result clearly indicates that indeed the properties of MRI turbulence in the 
unstratified shearing box model depend only on dimensionless wavenumber $kL$.
As a result, similar properties found in previous sections for tall boxes of $\lx = H$ are
also present in tall boxes with different $\lx$.

Returning to Figure~\ref{fig:alpha_aspect4}, the stress measured in our
simulations deviates slightly from the $(\lx/H)^2$ 
power law scaling at $\alpha \gtrsim 0.1$, which occurs in the largest boxes.
We speculate this is due to compressibility effects
\citep{sano2004}.  Such large values of $\alpha$ are associated with very strong
(nearly sonic) turbulence.  For example,
the rms density fluctuation for runs with large $\lx$ become as large as 
$\langle(\delta\rho/\rho_0)^2\rangle^{1/2} \sim 0.25$ in x2y8z8r64.  Compressibility
strongly damps the turbulence, and
prevents the turbulent magnetic field from growing even stronger.  Obviously 
compressibility effects must play a role at some point, as turbulence alone cannot
generate $\alpha \gg 1$.

\subsection{Explicit dissipation}
\label{sec:diss}
Previous work has shown that if explicit dissipation is included 
in the unstratified shearing box model with no net flux using the standard box size
$\lz/\lx=1$, then a converged value for the magnetic stress can be achieved
\citep[][]{FPLH2007}. 
The saturation level of the stress in this case is highly sensitive to the magnetic Prandtl
number $\rm{Pm}$.  It increases almost linearly with $\rm {Pm}$ in the range $ 4 \leq \rm{Pm} \leq 16$; 
while turbulence is completely suppressed once the magnetic Prandtl number is below some critical
value $\rm{Pm}_c \sim 2$-$4$ \citep{FPLH2007,SH2009}. Further study suggests this
behavior
shares similar origin to super-transient behavior in chaotic systems, and that the
lifetime
of the turbulent active
phase of the MRI increases exponentially with magnetic Reynolds number for fixed kinetic Reynolds number
\citep{Rempeletal2010,Riolsetal2013}.
In addition, studies with net flux and explicit dissipation have
also shown a somewhat weaker dependence on the magnetic Prandtl number \citep[][]{LL2007,SH2009},
with saturation levels appearing to reach asymptotic values in the $\rm{Pm}\ll 1$ limit \citep{Meheutetal2015}.  
These results have been used to argue it may be necessary to
include explicit dissipation in all simulations of the nonlinear regime of the MRI,
even at very large magnetic Reynolds number and
when the magnetic Prandtl number $\rm{Pm} \sim 1$.

However, we have shown above that for large aspect ratios ($\lz/\lx \gtrsim 1$), converged values
of the stress are achieved in ideal MHD.  To test whether the saturated stress is affected
by the inclusion of explicit dissipation, we have repeated simulations
in a tall box $(\lx,\ly,\lz) = (1,4,4)\,H$ but with four different combinations of kinematic
viscosity and Ohmic resistivity that give $(\rm{Re},\rm{Pm}) = (1600,7.8125)$ for run
x1y4z4r128pm8, $(3125,4)$ for x1y4z4r128pm4, $(3125,2)$ for x1y4z4r128pm2, and $(3125,1)$
for x1y4z4r128pm1, with resolution of $128/H$. Parameters are chosen to match previous
simulations in \citet{FPLH2007,SH2009}.
A resolution of $128/H$ ensures that the small scale dissipation is well resolved in
the parameter space we explored \citep{SH2009}. The results of all four runs are listed in Table~
\ref{tab:tab1}. We note all runs are simulated with the same initial conditions
as described in Section~\ref{sec:ic} except x1y4z4r128pm8, which is restarted from $t=50$ orbits of
run x1y4z4r128pm2 to reduce the computational cost.

\begin{figure*}
	\includegraphics[width=16cm]{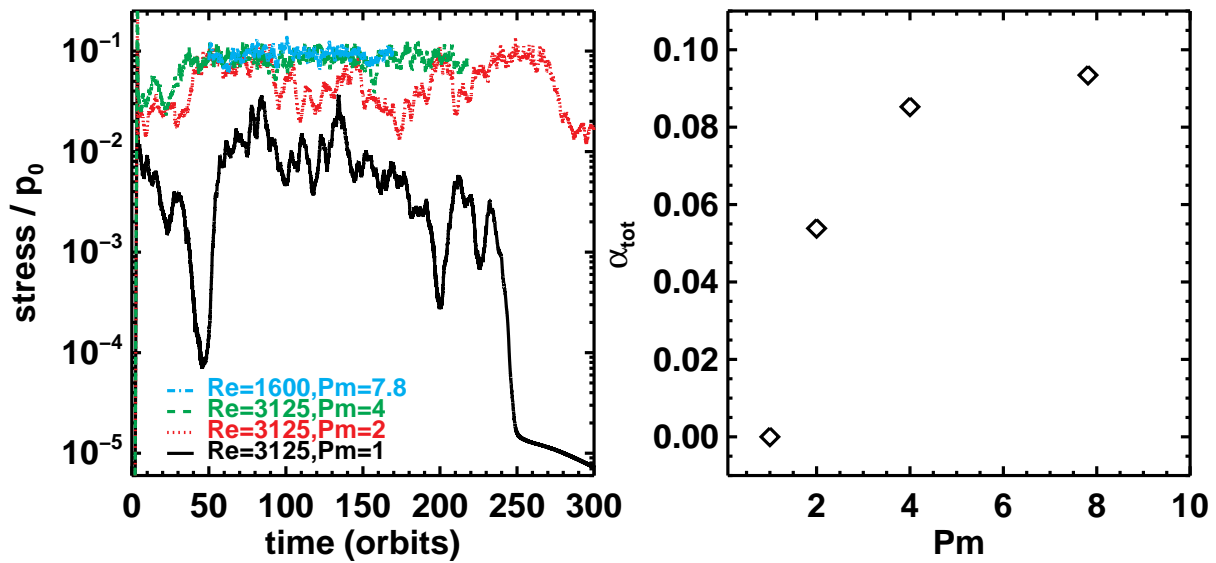}
\caption{\small{ Left: Time evolution of volume averaged total stress for
different magnetic Prandtl number $\rm{Pm}$; Right: Time averaged total stress-to-pressure ratio as
a function of 
$\rm{Pm}$. All curves are calculated using resolution $128/H$. We set $\alpha_{\rm tot}=0$ for
$\rm{Pm}=1$ case(run x1y4z4r128pm1), as turbulent transport vanishes after $250$ orbits. 
The saturated stress level is sensitive to the magnetic Prandtl number when $\rm{Pm}\lesssim 4$;
however this dependence becomes less significant when $\rm{Pm} \gtrsim 4$, unlike the linear
dependence found in standard boxes \citep[][]{FPLH2007,SH2009}. 
}	}
\label{fig:alpha_pm}
\end{figure*}

\begin{figure*}
	\includegraphics[width=8cm]{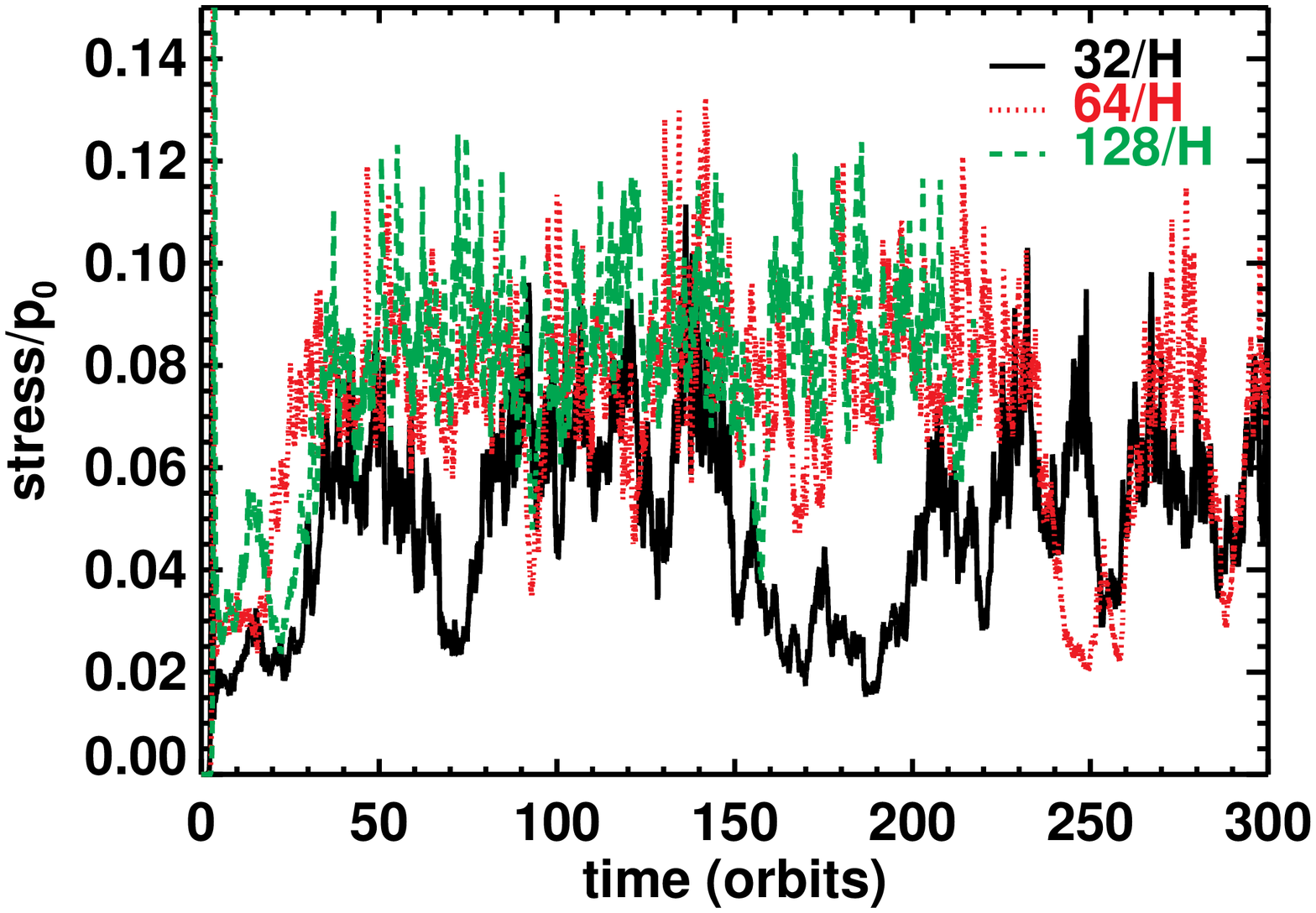}
	\includegraphics[width=8cm]{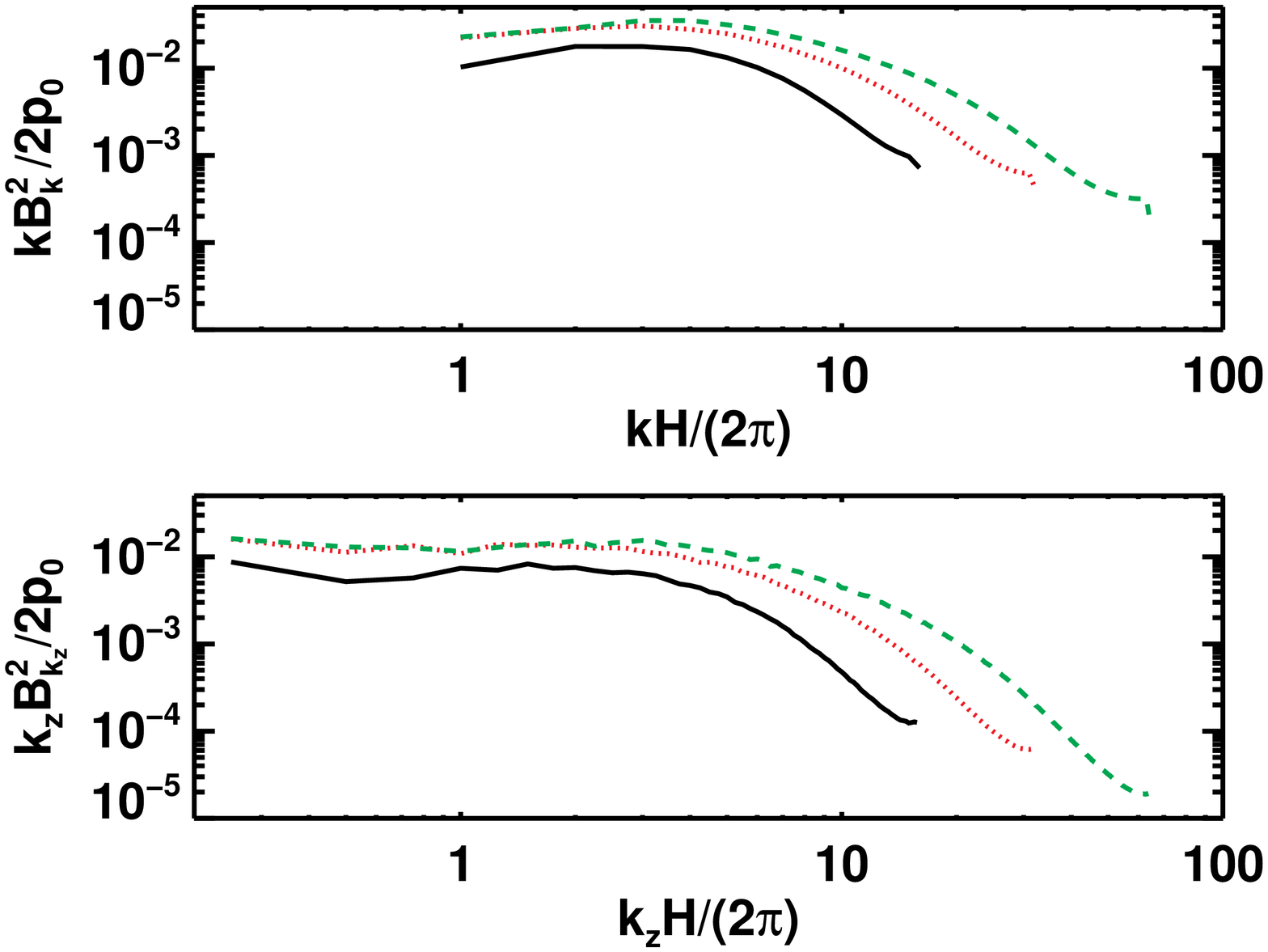}
\caption{\small{Similar to Figure~\ref{fig:convergence} and \ref{fig:convergence_pwr} but with explicit dissipation:$\rm{Re}
=3125$ and $\rm{Rm} =12500$. The results are very similar to the runs without explicit dissipation
in Figure~\ref{fig:convergence} and \ref{fig:convergence_pwr}. }}
\label{fig:convergence_resist}
\end{figure*}

We find for $\rm{Pm} \gtrsim 4$ (run x1y4z4r128pm4 and x1y4z4r128pm8), sustained stress and magnetic
energy are achieved in tall boxes. Moreover, the saturated stress and magnetic energy
for these runs are not much different from those without explicit dissipation. The time and volume averaged
total stress is $\simeq 0.085$ for run x1y4z4r128pm4, and $0.092$ for x1y4z4r128pm8, comparing to
$0.084$ for run x1y4z4r128 without explicit dissipation (see Figure~\ref{fig:alpha_pm} and
Table~\ref{tab:tab1}).  We find, in Figure~\ref{fig:convergence_resist}, both the volume averaged
stresses and magnetic energy spectra are quite similar to the tall box runs without explicit
dissipation as shown in right columns of Figure~\ref{fig:convergence} and \ref{fig:convergence_pwr}.  
It would seem that for taller boxes, in which
the nonlinear regime and saturation is controlled by dynamo action at least in the $\rm{Pm}\gtrsim
4$ regime, including explicit dissipation has little effect on the results. 
We note that the saturated stress, $\alpha_{\rm tot} \sim 0.085$ owning to the large scale mean field, t
is much greater than the values reported in run SZRe3125Pm4 of \citet[][]{SH2009}($\sim 0.013$), and
in run 128Re3125Pm4 of \citet[][]{FPLH2007} ($\sim 0.009$). 
We also note that the shape of the power spectrum is different than the standard short box
	simulations with explicit dissipation \citep[][]{Fromang2010} as the large scale dynamo
	observed in our tall box runs is completely
absent in this latter case. 

When $\rm{Pm}=2$ (run x1y4z4r128pm2), the stress in Figure~\ref{fig:alpha_pm} appears to be more
bursty compared to higher Prandtl number runs and those without explicit dissipation. The large
fluctuations might be a result of competition between super-transient decay and dynamo
growth \citep{TS2015}. Time averaging over the last $200$ orbits, we find $\alpha_{\rm tot} \sim 0.05$,
almost twice smaller than high $\rm{Pm}$ runs. The same parameters have also been explored previously
using a standard box, see run 128{\rm Re}3125{\rm Pm}2 in
Figure~8 of \citet[][]{FPLH2007} and SZRe3125Pm2 in Table~1 of \citet[][]{SH2009}, in contrast to our
tall box run x1y4z4r128pm2, no sustained MHD turbulence is observed in either
of these cases. 

The only run in which we find turbulence eventually decays is $\rm{Pm}=1$ (run x1y4z4r128pm1 in
Figure~\ref{fig:alpha_pm}).
After about $250$ orbits, both the Maxwell stress and magnetic energy abruptly drop by several orders of
magnitude and the flow becomes laminar. This finding confirms the existence of a critical Prandtl
number $\rm{Pm}_c$ for a given kinetic Reynolds number even when the box is tall ($\lz/\lx > 1$ and
$\ly/\lx = 4$). Together with the results of $\rm{Pm}=2$ run, it indicates a smaller critical value,
$1 \lesssim \rm{Pm}_c \lesssim 2$, comparing to $\rm{Pm}_c=2$-$4$ found in standard boxes
\citep{FPLH2007,SH2009}. As we mainly varies the $\rm{Re_M}$, we caution the readers that 
the results might possibly reflect the dependence of saturation amplitude of the MRI on another
dimensionless number, the Lundquist number $\rm{Lu}\equiv v_{\rm Az}^2/\eta\Omega$
\citep{sano2002}. We find the time averaged $\rm{Lu} \simeq 
332$, $105$ and $11$ for run x1y4z4r128pm4, x1y4z4r128pm2 and x1y4z4r128pm1
\footnote{ Take time average over turbulent active phase $t< 250$ orbits.}, getting closer to the
critical values reported in previous studies with vertical net flux
\citep{LL2007,tsd2007,pessahetal2007,MS2008}.

The result of \citet[][see their Figure~5]{Rempeletal2010}, based on the
statistics of MRI turbulence in a
standard $(\lx:\ly:\lz) = (1:\pi:1)$ box, would predict a
characterisic decay time $\sim 10$ shear time units for $\rm{Pm}=1$ and $2$ with
the same $\rm{Re}=3125$ as used here. However, we find sustained turbulence that
lasts more than several thousand
shear time units in our $\rm{Pm}=2$ run; and the active time for our $\rm{Pm}=1$ run is $\sim
200$ orbits, or $2000$ shear time units, much longer than the prediction. Moreover, 
\citet{FPLH2007} find, when changing $\rm{Re}$ from $12500$ to $25000$, the turbulence lifetime of
$\rm{Pm}=1$ increases from  $\lesssim 40$ orbits to $\sim 100$ orbits. This would suggest a
decaying time even shorter than $40$ orbits for $\rm{Re}=3125$ case, therefore much smaller than our result. 
Clearly, the aspect ratio of the computational domain may play a big role in
determining the dynamical lifetime of MRI turbulence \citep{Riolsetal2015}
in the unstratified, no net flux shearing box.

To sum up, we find the Prandtl number dependence of the turbulent stress in our tall box runs is
very different from that found earlier using a standard box (see right panel of
Figure~\ref{fig:alpha_pm}). When $\rm{Pm}\ge 4$, the saturated stress level is insensitive to the
inclusion of explicit dissipation. We find converged turbulent stress for runs with and without
explicit dissipation. The saturated stress shows stronger dependence when $\rm{Pm}<4$ as it drops
linearly from $\rm{Pm}=4$ to $2$. By choosing a taller box which promotes dynamo
action, we find the dynamical lifetime of MHD turbulence is extended, and the 
critical magnetic Prandtl number is reduced
compared to the standard smaller box simulations.

\section{MODELING THE DYNAMO}
\label{sec:dynamo}
\subsection{Dynamo cycle period}
\label{sec:cycle}
\begin{figure}
	\includegraphics[width=\columnwidth]{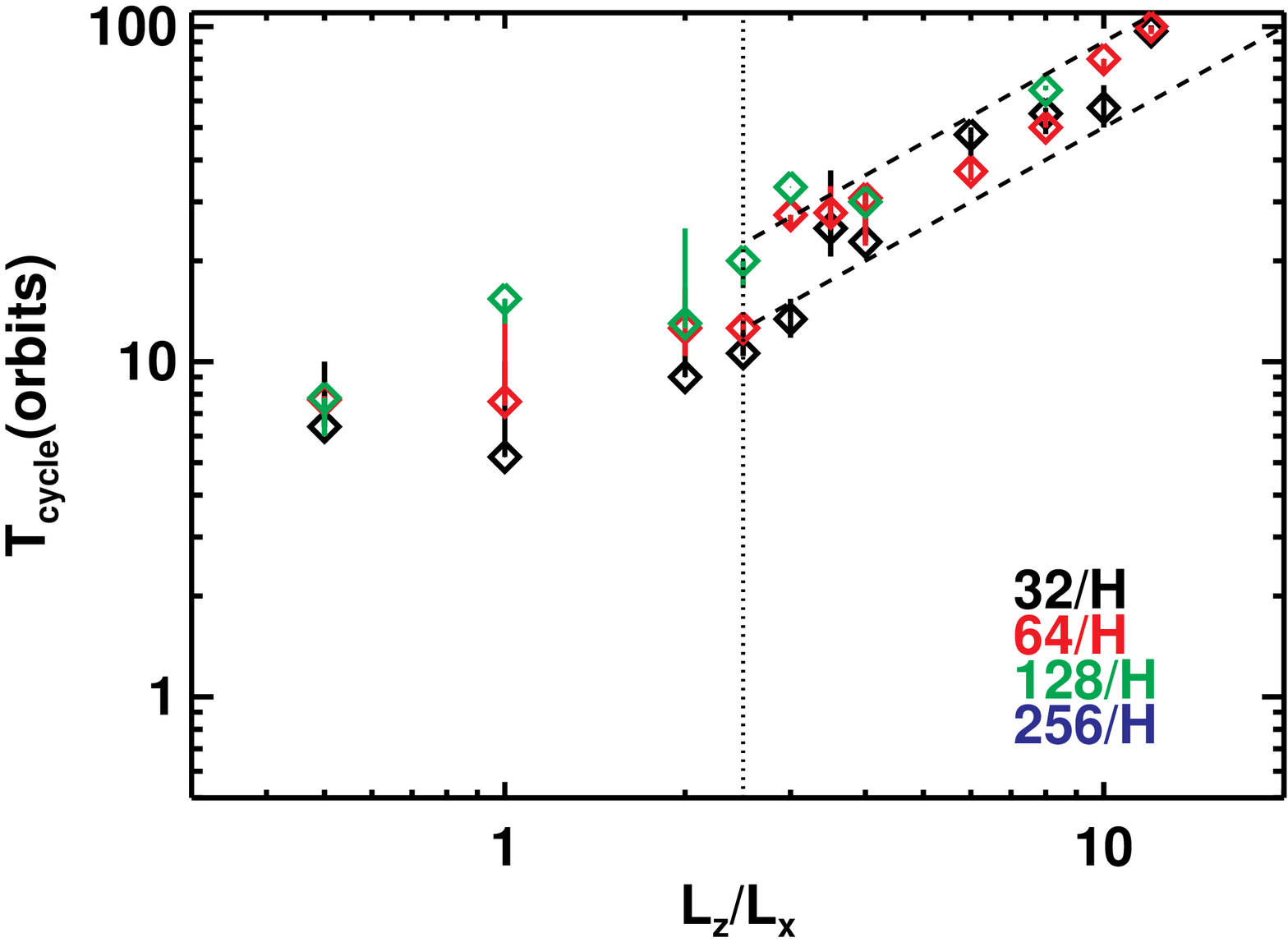} \\
	\includegraphics[width=\columnwidth]{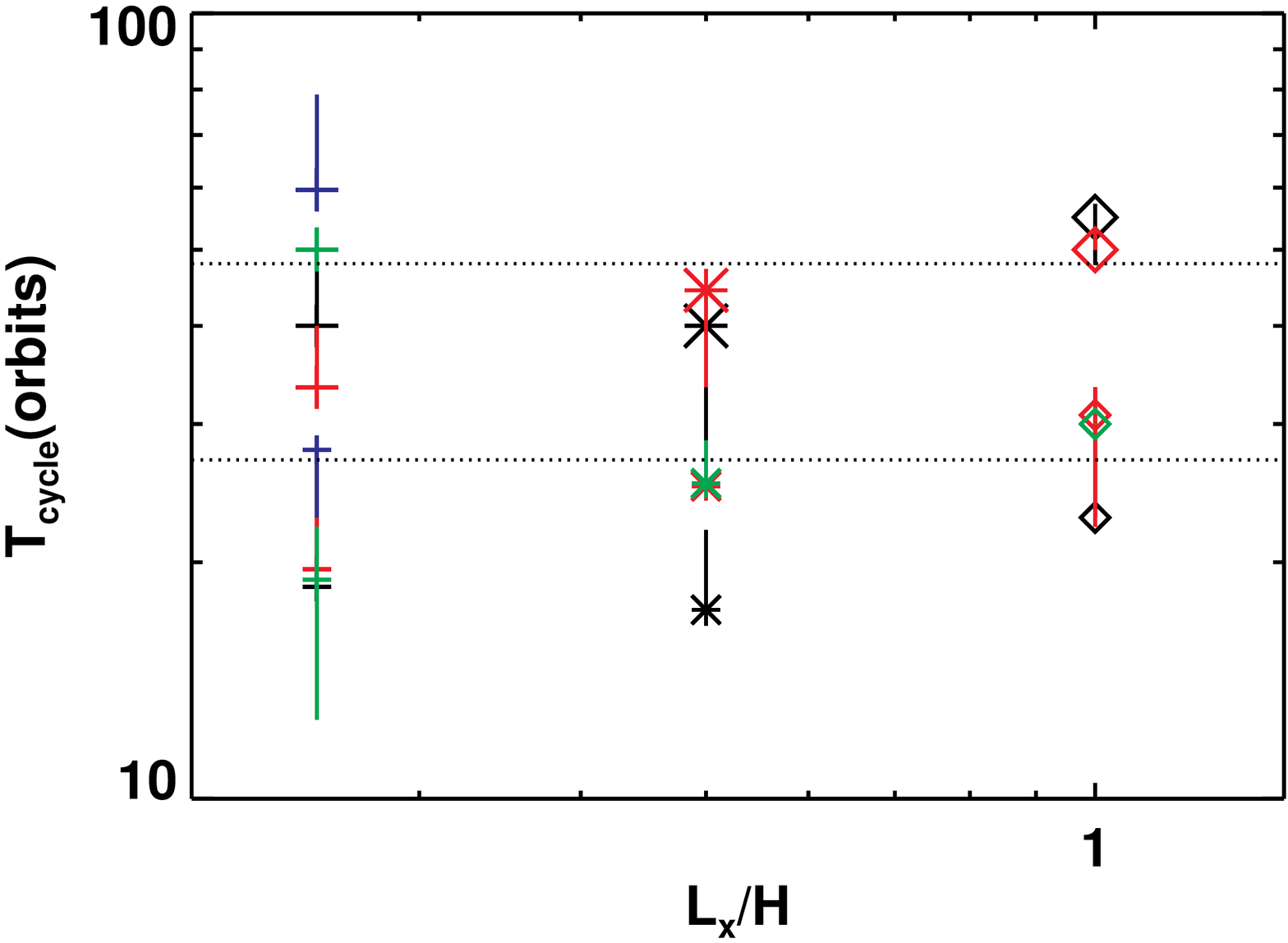} \\
	\includegraphics[width=\columnwidth]{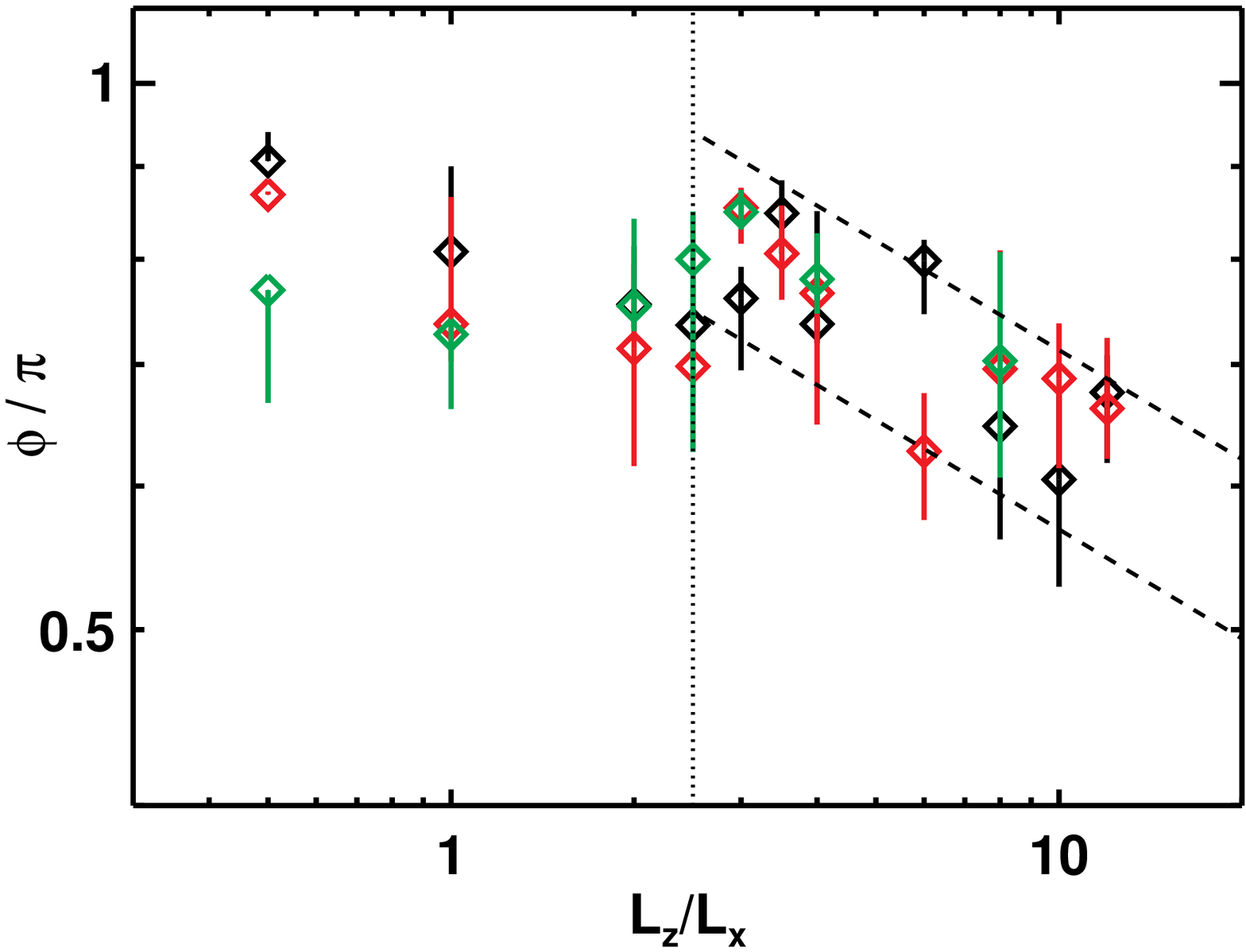}
\caption{\small{
Top: Dynamo cycle periods for different $\lz/\lx$ with fixed $\lx/H=1$, dashed lines
show $\propto \lz/\lx$ power law. 
Middle: Dynamo periods vs. box sizes (small symbols for $\lz/\lx=4$, and larger symbols for $\lz/\lx=8$). 
The dotted horizontal lines at $\Tcycle=27$ and $48$ orbits show the constancy of cycle period with
respect to various box sizes. 
Bottom: Phase lag between radial and azimuthal mean field as calculated in
Equation~\ref{eq:phase_lag}, dashed lines follow $\propto (\lz/\lx)^{-0.2}$ scaling.
The dotted vertical lines at $\lz/\lx=2.5$ separate the standard and tall boxes. 
The error bars of $\Tcycle$ are the variations measured with two independent time sequences; the error
bars of phase lag show the max/min values given the variations of the measured $\Tcycle$ and time
delay $\Delta t$.}}
\label{fig:tcyl_aspect}
\end{figure}
Our previous results indicate that shearing box simulations of the MRI with a large
aspect ratio, dynamo action produces strong ordered toroidal fields on vertical
scales $\lx$.  In this section we explore models that might explain this dynamo action.

An important property of the dynamo is that at any given vertical location $z$, the
toroidal magnetic field is cyclic, and the period of these
cycles is an important clue to the mechanism of dynamo action.  As illustrated
in Figure~\ref{fig:zt_cycle}, cyclic patterns of the mean field $\ob{B}_y$ becomes strong and
regular for those boxes with $\lz \gtrsim 8\,H$. It is relatively easy to identify the cycle period from
the space-time diagrams. 
However, for smaller boxes, e.g., $\lz = 4\,H$ as shown in
Figure~\ref{fig:zt_cycle}, the cycle is less regular and it is more difficult to extract a single value of
the period $\Tcycle$. 
Thus, we measure the cycle period based on the power spectrum density (PSD) of the largest vertical mode
($k_{\rm z}=2\pi/\lz$) of $\ob{B}_y (t)$ in Fourier space, i.e.,
\beq
{\rm PSD~~of~~}\wt{B}_y(t) ,~~~{\rm where~~~} \wt{B}_y (t) \equiv \int\!\ob{B}_y(z;t) e^{-i k_{\rm
z} z} dz \,.
\label{eq:tcyl_def}
\enq
We can also estimate errors from the range of $\Tcycle$ as measured from
two independent time sequences. 
We have applied this
measurement to all simulations regardless of the box size, but note the physical meaning of
$\Tcycle$ in simulations with a small aspect ratio ($\lz/\lx < 2$) is less clear due to the
absence of dynamo action.
We list all values of $\Tcycle$ measured in this way in Table~\ref{tab:tab1} for reference. 

{Once dynamo is at action, there is in general a phase lag between the radial mean field $\ob{B}_x$ and the
azimuthal $\ob{B}_y$. In order to obtain the phase shift, we first apply filter on
our mean field and get the largest vertical mode $\ob{\mathbf{B}}^{(1)}(z,t)$, where `$(1)$' denotes
$k_z = 2\pi/\lz$ only. We then calculate the cross-correlation between two time sequences, $\ob{B}_x^{(1)}$ and
$\ob{B}_y^{(1)}$ at given $z$. The exact time delay $\Delta t$ between radial and azimuthal mean
field are the lag which gives the most negative value of correlation. In Table~\ref{tab:tab1}, we
list the vertically averaged $\Delta t$, and the errors from the variation of $\Delta t$ at
different height. The azimuthal mean field therefore lags behind the radial field by 
\beq
\phi = \pi - 2\pi \frac{\Delta t}{\Tcycle} 
\label{eq:phase_lag}
\enq
in phase.
}

In the top panel of Figure~\ref{fig:tcyl_aspect}, we plot the measured $\Tcycle$ for all $\lx=1\,H$
runs.  In general, we find the cycle period is much longer than the dynamical (orbital) time,
ranging from several to tens of orbital periods. The measured $\Tcycle$ becomes longer as $\lz/\lx$ 
is increases with a slope slightly shallower than unity.
In contrast to the aspect ratio dependence, $\Tcycle$ is not very sensitive to the box size.
As shown in the middle panel of Figure~\ref{fig:tcyl_aspect}, $\Tcycle$ for $\lz/\lx = 4$
scatters between $\sim 20$ and $30$ orbits, while it varies around
$\sim 50$ orbits for $\lz/\lx = 8$ with
$\lesssim 50\%$ fluctuations.
{The phase lag between radial and azimuthal mean field does not depend on box size either (see
Table~\ref{tab:tab1}). It only weakly depends on the aspect ratio, roughly $\propto(\lz/\lx)^{-0.2}$
as shown in the bottom panel of Figure~\ref{fig:tcyl_aspect}.}

\subsection{A toy model}
\label{sec:toymodel}
Following the model proposed in \citet[][]{LO08}, we also try to fit the dynamo cycles observed
in our simulations with the following nonlinear model:
\beq
\partial_t\Bxk = \gamma\Byk (t-t_r)\frac{\left| \Byk (t-t_r)\right|-B_r}{B_r}\,,
\label{eq:toymodel1}
\enq
\beq
\partial_t\Byk = -q\Omega\Bxk - \beta \Byk(t-t_r)\,,
\label{eq:toymodel2}
\enq
where $\Bxk$ and $\Byk$ are horizontally averaged radial and azimuthal
field that are filtered to conserve only the largest vertical mode ($k_z=2\pi/\lz$, represented with
the superscript `$(1)$'). 
Here $t_r$ characterizes the time delay between the EMF and large scale magnetic field. When the
mean field $|\Byk|$ is smaller than $B_r$, the $\gamma$ term effectively amplifies the
$\Bxk$ field, which in turn lead to $\Byk$ growth via shear.  In the 
opposite case, when $|\Byk|$
exceeds $B_r$, the $\gamma$ term starts to damp the radial field, and the
azimuthal field is thus
reduced via the $\beta$ term. 
In the fitting, we choose $t_r= 2/|q\Omega|=4\Omega^{-1}/3$ as
suggested by linear analysis \citep[][]{LO08-2,LO08}, and 
$B_r/(\sqrt{4\pi\rho_0}\cs)=0.3$, a factor $\sim 2.5$ greater than used in \citep[][]{LO08} to
better fit the mean field amplitude observed in our simulations. We then
vary $\gamma$ and $\beta$ to match the observed cycle period and long-term amplitude.
Empirically, $\gamma$ determines the cycle frequency, and $\beta$ prevent the solution from
diverging.

\begin{figure}
	\includegraphics[width=\columnwidth]{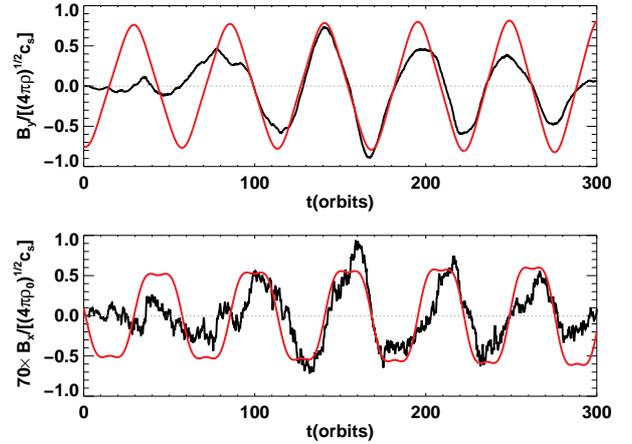}
\caption{\small{A toy model fit (red, based on Equation~\ref{eq:toymodel1} and \ref{eq:toymodel2}) to our
unstratified x1y4z8 run (black). The top panel shows the largest vertical mode of mean
$\ob{B}_y$ at $z=-3\,H$; the bottom panel shows the corresponding radial field. Dotted lines in
both panels mark zero amplitude level. }}
\label{fig:lomodel_unstratified}
\end{figure}
We focus on fitting the results for run x1y4z8r32 as it exhibits a well organized
cyclic pattern. Without loss of generality, we fit the filtered $\Bxk$ and $\Byk$ at
$z=-3\,H$. With $\gamma=2\times10^{-4}$ and $\beta=4\times10^{-4}$, the model is shown as
the red
curves in Figure~\ref{fig:lomodel_unstratified}. 
For times after $100$ orbits, the model provides a
reasonable fit to the spatial filtered simulation data.
{
The model predicts the
correct dynamo period and relative strength between $\Bxk$ and $\Byk$. However the
phase shift between $\Bxk$ and $\Byk$ is $\pi/2$, slightly smaller than the simulation data, which
is $\phi \simeq (0.65\pm 0.08)\pi$. This phase lag is also different than what is found in
\citet{Heraultetal2011} ($\sim \pi$), but is close to $3\pi/4$, the phase shift of a marginally excited dynamo
wave in a $\alpha$-$\Omega$ dynamo \citep{BS2005}.
}
As the values of $B_r$ and $t_r$ are uncertain, there
are likely to be degeneracies in the model parameters.

\begin{figure*}
	\includegraphics[width=16cm]{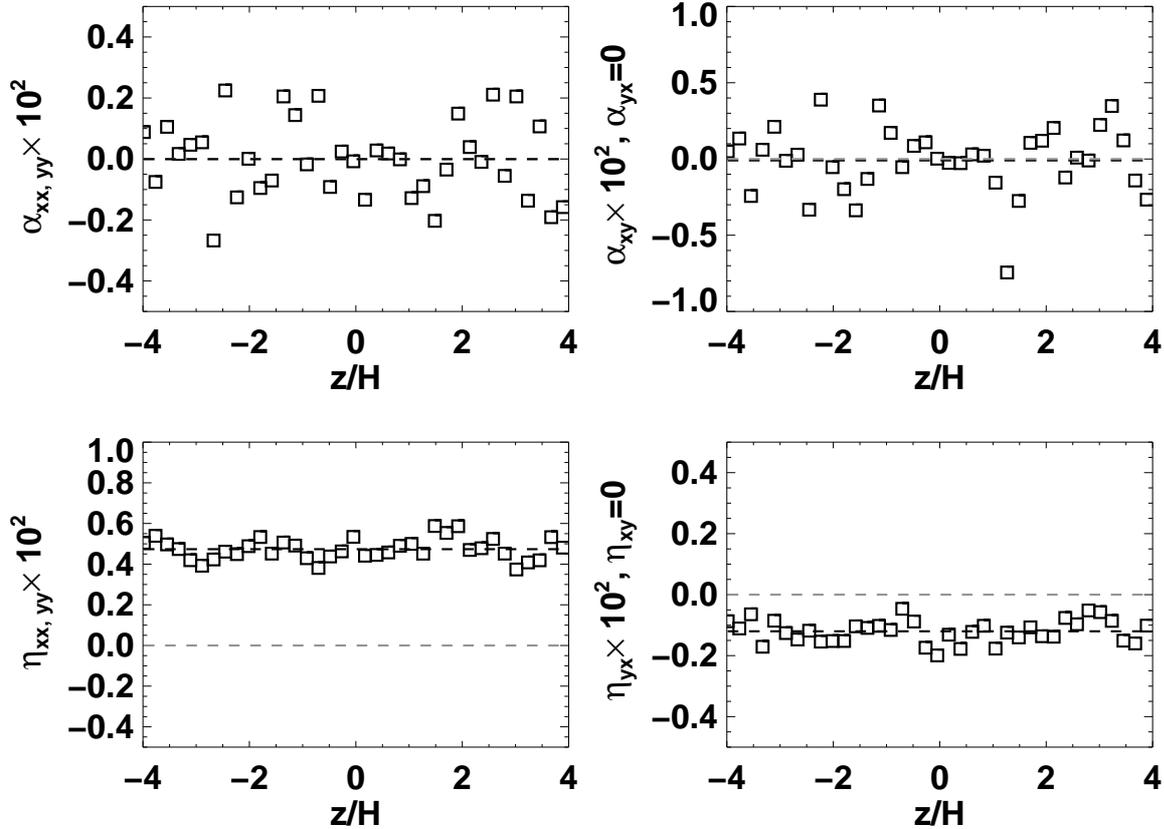}
\caption{\small{Turbulent transport coefficients measured via simulation at $t=10$-$50$ orbits,
		assuming closure model like Equation~\ref{eq:closure}. We further set
		$\alpha_{xx}=\alpha_{yy}$, $\alpha_{yx}=0$, $\eta_{xx}=\eta_{yy}$ and $\eta_{xy}=0$
		in this fitting. Square symbols are the time averaged coefficients at evenly spaced
		disk height. The black dashed lines show the vertical averages and the gray dashed
		lines are set to zero to help read this plot. 
		Clearly, negative $\eta_{yx}$ is obtained for all $z$ indicating an
anti-diffusive growth of magnetic field.}}
\label{fig:etayx}
\end{figure*}
Recent work has explored this model as a magnetic analogue to the `shear current' effect proposed by
\citet[][]{RogaKlee2004} in which the non-isotropic contribution of the turbulent diffusion can drive
large scale magnetic field growth in non-helical turbulent shearing flows. 
Interestingly,  \citet[][]{squire2015a} recently find negative off-diagonal turbulent diffusivity
(a negative $\eta_{yx}$ term which drives exponential growth of $B_x$, see their Equation~(2) and
(3), or Equation~\ref{eq:closure} in this paper for its
definition) by direct measurement in simulations assuming a simple closure model,
\beq
{\mathcal{E}}_i = \alpha_{ij}\ob{B}_j - \eta_{ij}\epsilon_{jkl}\partial_k\ob{B}_l\,, ~~~~
i,j,l \in\{x,y\}\,, k=z
\label{eq:closure}
\enq
where the turbulent EMF, $\mathcal{E}\equiv\ob{v\times b}$, and mean field $\ob{\mathbf{B}}$ can be
measured from the simulation data, and $\epsilon_{jkl}$ is the Levi-Civita permutation
symbol. After setting $\alpha_{xx}=\alpha_{yy}$, $\alpha_{yx}=0$, $\eta_{xx}=\eta_{yy}$ and
$\eta_{xy}=0$, they find an average $\eta_{yx}\sim - 10^{-4}H^2\Omega$. Following their
procedure, i.e. computing $\langle \mathcal{E}_i M\rangle$ from Equation~\ref{eq:closure} for each of
$M = (\ob{B}_x, \ob{B}_y, \partial_z\ob{B}_x, \partial_z\ob{B}_y)$ and solving the resulting matrix
equations at each time point, we also find negative $\eta_{yx}\simeq -1.6\times 10^{-3} H^2\Omega$
by time averaging over the dynamo's growth phase $t=10$-$50$ orbits in x1y4z8r32.

Alternatively, we can also solve Equation~\ref{eq:closure} at any given $z$ directly assuming
time-independent transport coefficients (but a function of height) with multiple linear regression
method. The results are shown in Figure~\ref{fig:etayx}. All $\alpha$'s fluctuate about zero and are
close to zero when averaging over $z$. 
The diagonal $\eta_{xx} \sim 5 \times 10^{-3} H^2\Omega$ are dissipations of magnetic field. The non-diagonal
term $\eta_{yx}\sim 10^{-3} H^2\Omega$ for all $z$ is consistent with that of \citet[][]{squire2015a},
indicates a dynamo growth. We find this negative $\eta_{yx}$ behavior does not change with non-zero
$\alpha$'s and
$\eta$'s, or unequal $\alpha_{xx}$ and $\alpha_{yy}$, as long as the diagonal $\eta$'s are kept to
be the same ($\eta_{xx}=\eta_{yy}$).  We also obtain similar values of $\eta_{yx}$ in the well established dynamo
cyclic phase using data at $t=100$-$200$ orbits. 

However, comparing a time dependent $\eta_{yx}$ throughout a dynamo cycle
\footnote{{{It is calculated again by fitting moment equations of the closure~\ref{eq:closure}
	at each time point (without time average) using $\mathcal{E}$ and $\mathbf{B}$ values
	directly from the simulation. We can also fit Equation~\ref{eq:closure} with
	filtered data, e.g., by keeping only the first few wavenumbers. The resulting $\eta_{\rm
yx}$ can switch sign, however is so noisy that we can hardly retrieve any meaningful results from it.}}}
shows no evidence for sign change when
the mean field $|\ob{B}_y|$ surpasses some critical magnetic field strength as designed in the toy
model, although we do see the filtered EMF $\mathcal{E}_y^{(1)}$ changes sign over a cycle {
with the projection method used in \citet[][see their Eqaution~(12,13)]{LO08} and after some
smoothing.} Further study is still required to explain the inconsistency we find here, 
and identify the dynamo mechanisms ultimately, but is beyond the scope of this paper.

\subsection{Connection to the stratified shearing box}
It is interesting to compare the dynamo cycles in our unstratified shearing box
simulations with those observed in stratified disks. Simulations using the
stratified shearing box
always find strong dynamo cycles in the toroidal field
\citep[e.g.,][]{Brandenburgetal1995,shgb96,ZR2000,Gressel2010,OM2011,kapylaetal2013}. 
As the unstratified shearing box disks approximates the
midplane of a stratified disks, there may be some relation between the dynamo behavior seen in
both cases.

\begin{figure}
	\includegraphics[width=\columnwidth]{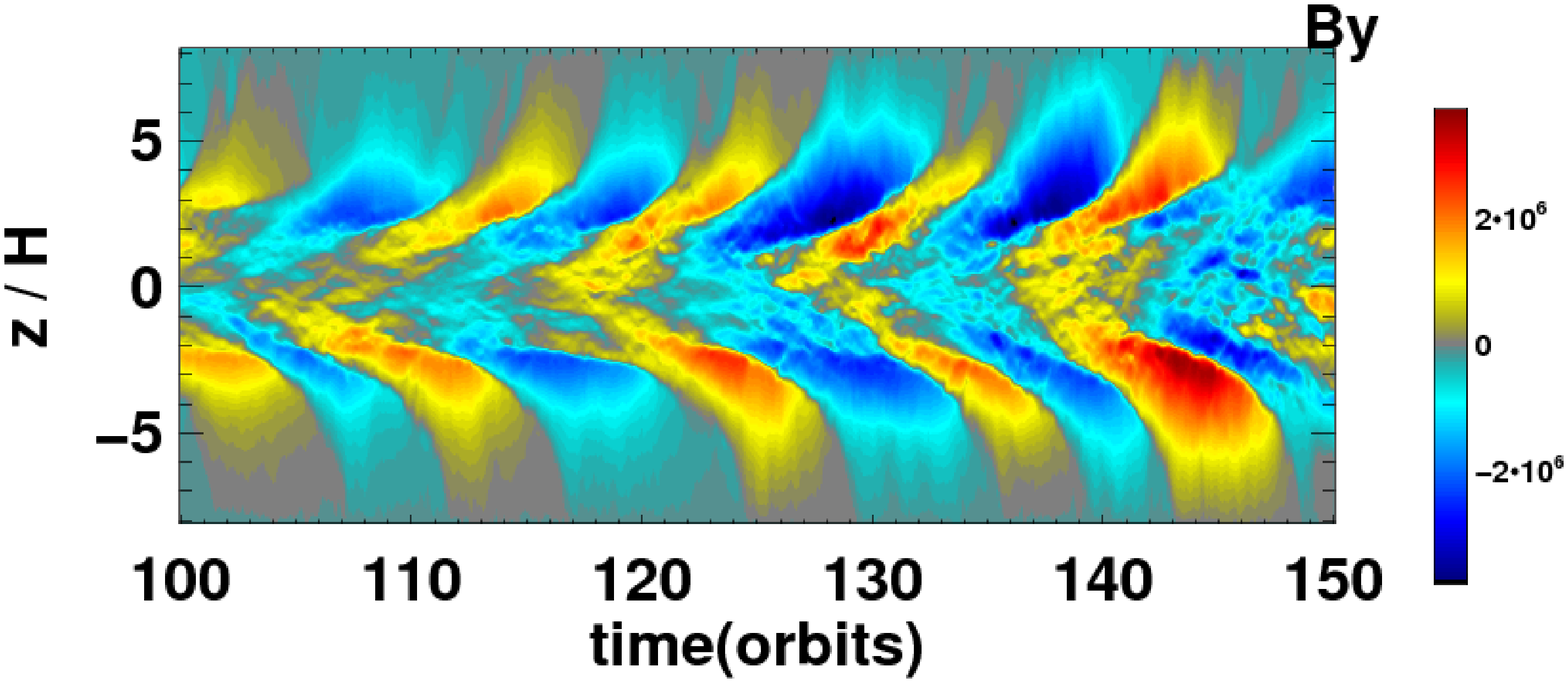}\\
	\includegraphics[width=\columnwidth]{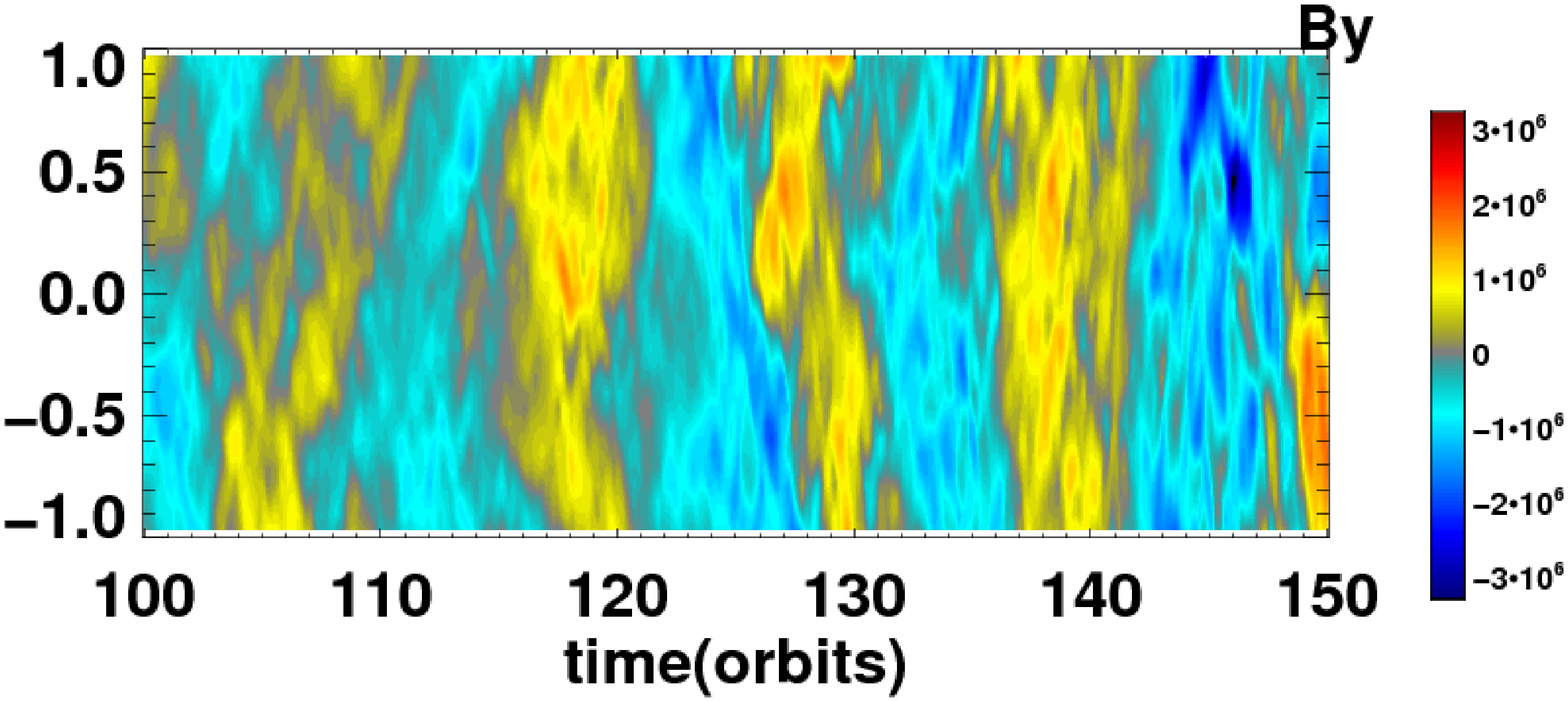}
\caption{\small{Top: Space-time diagram of $\ob{B}_y$ in a stratified box STD32 \citep[][]{Shi2010}.
Bottom: a blow-up view of the left to show the main body of the disk within $\pm$ H.}}
\label{fig:stratified}
\end{figure}
We start by analyzing an existing stratified simulation data (run STD32) first published
in \citet[][]{Shi2010}. Due to buoyancy, the mean magnetic field is expelled toward the disk surface producing a
regular pattern known as the butterfly diagram (see Figure~$6$ of \citet[][]{Shi2010})
reproduced here in Figure~\ref{fig:stratified}. By extracting data within $\pm H$ of the
midplane,
we find a clear dynamo cycle with a period $\sim 10$
orbits as shown in the bottom panel of Figure~\ref{fig:stratified}.  At any given time, the mean
azimuthal field is roughly uniform in $z$, and has one sign throughout the subvolume near
the midplane (to preserve the constraint of no net flux, field of the opposite sign
is located above and below the midplane).

\begin{figure}
	\includegraphics[width=\columnwidth]{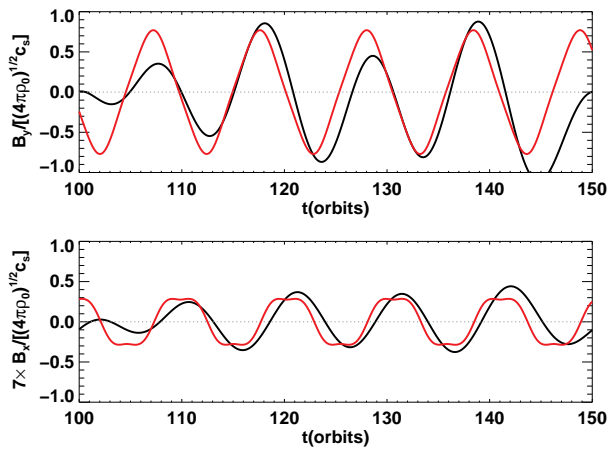}
\caption{\small{ A toy model fit (based on Equation 11 and 12) to the stratified disk STD32
		(see Figure~\ref{fig:stratified}; \citet[][]{Shi2010}). The fit is in red, the
		vertically averaged mean $B_y$ (top) and $B_x$ (bottom) within $\pm$ H are in black.
		Dotted lines in both panels mark zero amplitude level. }}
\label{fig:lomodel_stratified}
\end{figure}
We can also model the dynamo in the stratified shearing box with the toy model described
in section~\ref{sec:toymodel}. Again, we adopt
$t_r = 4/(3\Omega)$ and $B_r/(\sqrt{4\pi\rho_0}\cs) = 0.3$, where $\rho_0$ and $\cs$ are the
midplane density and sound speed in the stratified disk. We fit the vertically averaged $\ob{B}_x$
and $\ob{B}_y$ with parameter values of $\gamma = 10^{-3}$ and $\beta = 0.014$ as shown in
Figure~\ref{fig:lomodel_stratified}. Similar to the unstratified case, the toy model captures the
cycle period and relative strength of the field components correctly, but slightly
underestimates the relative phase between
$\ob{B}_x$ and $\ob{B}_y$ ($\phi/\pi\simeq 0.6^{+0.1}_{-0.2}$).

The overall similarity between the dynamo in the unstratified and stratified disks suggests the same 
mechanism could act in both cases. We speculate that the core of the stratified disk resembles our
tall box runs. Extrapolating the linear scaling found in the left panel of
	Figure~\ref{fig:tcyl_aspect} to $\lz/\lx \lesssim 2$ gives $\Tcycle\sim 10$ orbits which
matches the cycle period of the stratified shearing box very well.

\begin{figure*}
	\includegraphics[width=16cm]{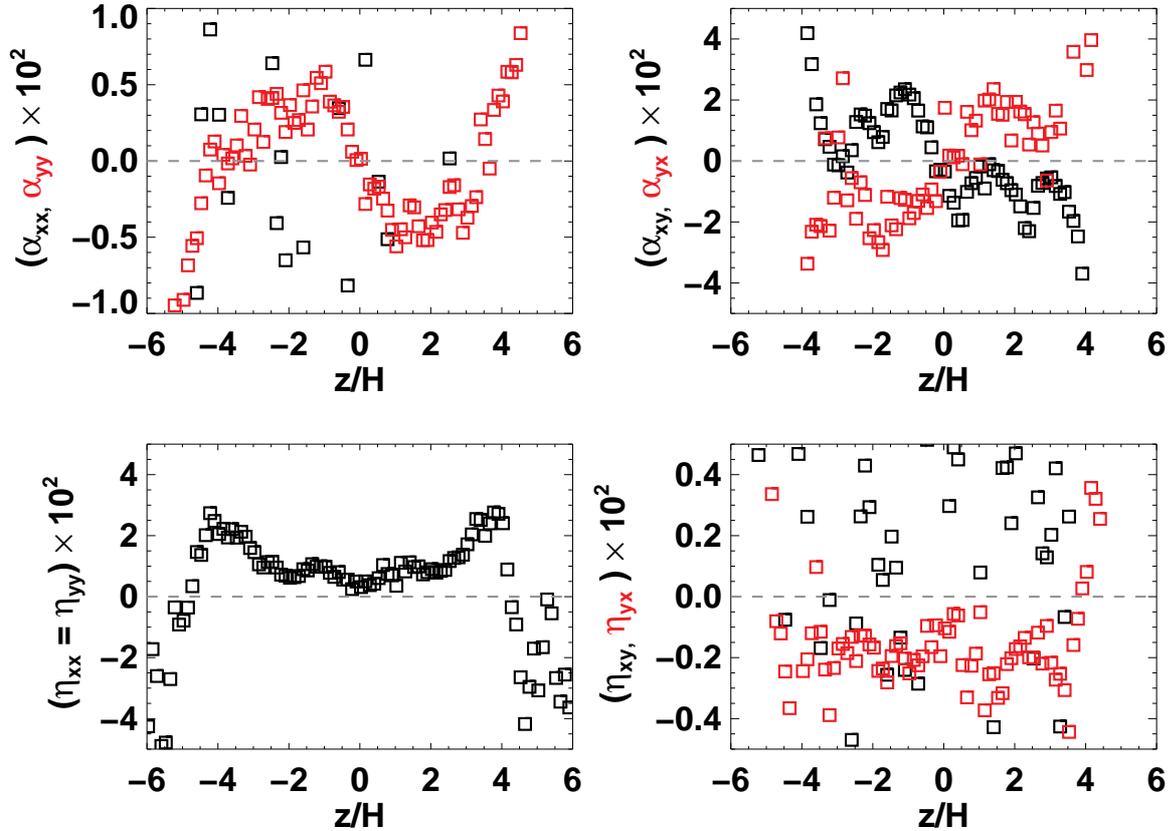}
\caption{\small{Similar to Figure~\ref{fig:etayx}, turbulent transport coefficients measured via
		stratified disk as in Figure~\ref{fig:stratified} are plotted assuming only
		$\eta_{xx}=\eta_{yy}$. In each panel, different colors represent different
		coefficients as labeled on its $y$-axis. We see similar
		negative $\eta_{yx}$ as in the unstratified disk within $\pm4\,H$, also a
		negative (positive) $\alpha_{yy}$ below (above) $4\,H$ in the upper half of the disk,
		antisymmetric with respect to the midplane.}}
\label{fig:std_etayx}
\end{figure*}
Assuming similar correlation between EMF and mean field as described in
Section~\ref{sec:toymodel} \citep[][]{squire2015a,SB2015b}, we find again a negative off-diagonal resistivity
$\eta_{yx} \sim -10^{-3}H^2\Omega$ (Figure~\ref{fig:std_etayx}), which favors an anti-diffusive
dynamo model similar to the unstratified case. 
However, we find, unlike the unstratified case, an $\alpha$-effect is also present in the
stratified box. The diagonal $\alpha_{yy}$ is antisymmetric with respect to the midplane. In the
upper half of the disk, it is negative below $4\,H$, and systematically positive above $4\,H$.
The negative $\alpha$-effect (within the upper half of the disk) is generally attributed to the
dynamo growth in previous stratified simulations and the negative sign is caused by the influence of
the field buoyancy and/or strong shear in accretion
disks \citep[e.g.,][]{Brandenburgetal1995,RP2000}. The positive
$\alpha$-effect has also been recently proposed to explain the dynamo
\citep[][]{Gressel2010,GP2015}. Together
with the $\eta$-effect, it seems to suggest a combination of direct and indirect dynamo mechanisms
\citep[][]{BT2004,Gressel2010} in stratified shearing box. Further investigations are required to
identify whether one or both mechanisms could exist and further cause the dynamo cycles. 

\section{CONCLUSIONS}
\label{sec:conclusion}
In this paper, we have studied MRI turbulence in the unstratified shearing box model
with no net flux, and in particular how the size and aspect ratio of the computational
domain affects the saturation amplitude and stress.  We have performed simulations both
with and without explicit dissipation.
Previous simulations of this particular shearing box model have shown that the
saturation amplitude of the MRI depends on numerical resolution.  
However, we have shown
that the geometry of the computational domain also strongly affects the result.
Our main results are summarized as follows: 
\begin{enumerate}
\item{The amplitude of the stress and turbulence driven by the MRI converge to values 
independent of numerical resolution if the aspect ratio of the computational domain
$\lz/\lx >2.5$, even without explicit dissipation.  These values are proportional to the box size
$\lx^2$. Only when the aspect ratio $\lz/\lx < 2$ do these levels depend on resolution.}
\item{When $\lx=H$ and $\lz/\lx >2.5$ the converged value of the saturated stress is greatly
increased compared to previous values reported for $\lz/\lx=1$, and can be as
large as $\alpha \gtrsim 0.1$.  These values are similar to those required
by observations of dwarf novae disks in outburst, perhaps negating some of the concerns
expressed in \citet[][]{kpl2007}. }
\item{The Fourier power spectra of runs with large aspect ratio are very different from those with
$\lz/\lx \sim 1$.  In particular, most of the power in the magnetic field is at small $k$,
independent of resolution.}
\item{For the limited range of parameter values explored in this paper, we
find the saturation level of the MRI is independent of whether explicit dissipation
(viscosity and resistivity) is included or not, provided that the magnetic Reynolds
number is large.  Turbulence can be suppressed at small ${\rm Re_M}$, corresponding
to small magnetic Prandtl number $\rm{Pm}\lesssim 4$,
however simulations utilizing a
taller box show a relatively smaller critical Prandtl number and much longer lifetime of turbulence than
those with $\lz/\lx \sim 1$.}
\item{Cycles of strong large scale toroidal magnetic field with alternating sign
on scales of a few $\lx$ are generated in domains with $\lz/\lx >2.5$ via a dynamo.
This large scale magnetic field sets and sustains the
turbulence independent of numerical resolution.}
\item{Some aspects of the cyclic dynamo can be modelled with anisotropic turbulent resistivity
\citep{LO08}, although the physical mechanisms driving the dynamo are unclear.  Similar dynamo
cycles observed in a stratified shearing box simulation suggest the stratified and
unstratified disks might share common dynamo mechanisms.}
\end{enumerate}

As discussed in the Introduction, the
unstratified zero-net-flux shearing box model is unlikely to have much relevance to real
astrophysical disks.  Instead, it is likely that different local ($L \sim H$) patches of the disk
are threaded by a broad range of net field strengths, even if the total field threading the
disk is zero.  Thus, global models of astrophysical disks are best constructed
from an ensemble of net flux shearing box simulations.  Nevertheless, studies
of the unstratified zero-net-flux shearing box are of interest in order to study MHD
dynamo action in a simple, well-posed model.  Moreover, we have shown that
some aspects of the dynamo observed in stratified shearing boxes may be related to 
properties of the dynamo discussed here.
Thus, further study of the MRI dynamo in domains with large aspect ratio, in both
compressible and incompressible MHD, and both with and without explicit dissipation
are warranted.

\section*{Acknowledgements}
We thank Steve Balbus, Eric Blackman, Charles Gammie, Geoffroy Lesur and Jonathan Squire for
thoughtful comments on an early draft of this paper. We also thank Amitava Bhattacharjee,
Fatima Ebrahimi, Sebastien Fromang and Geoffroy Lesur for encouraging discussions. 
We thank Shane Davis for sharing the code to compute the power spectrum. 
This work was supported in part by the
National Science Foundation under grant PHY-1144374, "A Max-Planck/Princeton Research Center for
Plasma Physics" and grant PHY-0821899, "Center for Magnetic Self-Organization".
Resources supporting this work were provided by the Princeton Institute of Computational Science
(PICSciE) and Engineering and Stampede at Texas Advanced Computing Center (TACC), The University of Texas at
Austin through XSEDE grant TG-AST130002.
\clearpage

{\renewcommand{\arraystretch}{1.3}
\begin{deluxetable}{lccccccccccc}
\tabletypesize{\footnotesize}
\tablecolumns{12} \tablewidth{0pc}
\tablecaption{Simulation parameters and results\label{tab:tab1}}
\setlength{\tabcolsep}{-0.03in}
\tablehead{
\colhead{\begin{tabular}{l} Name\tablenotemark{(a)}                  \\                   \end{tabular}} &
\colhead{\begin{tabular}{c} $\alpha_{\rm tot}$  \tablenotemark{(b)}  \\ $(\times 0.01)$   \end{tabular}} &
\colhead{\begin{tabular}{c} $\alpha_{\rm M}  $  \tablenotemark{(b)}  \\ $(\times 0.01)$   \end{tabular}} &
\colhead{\begin{tabular}{c} $\alpha_{\rm R}  $  \tablenotemark{(b)}  \\ $(\times 0.01)$   \end{tabular}} &
\colhead{\begin{tabular}{c} $\Delta T$\tablenotemark{(c)}            \\ $({\rm orbits})$  \end{tabular}} &
\colhead{\begin{tabular}{c} $\Delta T_{\rm avg}$\tablenotemark{(d)}  \\ $({\rm orbits})$  \end{tabular}} &
\colhead{\begin{tabular}{c} $T_{\rm cycle}$ \tablenotemark{(e)}      \\ $({\rm orbits})$  \end{tabular}} &
\colhead{\begin{tabular}{c} $\Delta t$ \tablenotemark{(f)}           \\ $({\rm orbits})$  \end{tabular}} &
\colhead{\begin{tabular}{c} $\dLangle\dfrac{B^2}{8\pi P_0}\dRangle_t^{(g)}$               \end{tabular}} &
\colhead{\begin{tabular}{c} $\dLangle\dfrac{\rho v^2}{2 P_0} \dRangle_t^{(g)}$            \end{tabular}} &
\colhead{\begin{tabular}{c} $\dLangle\dfrac{\ob{B}^2}{8\pi P_0}  \dRangle_t^{(g)}$        \end{tabular}} &
\colhead{\begin{tabular}{c} $\dLangle\dfrac{\ob{B}^2_y}{8\pi P_0}\dRangle_t^{(g)}$        \end{tabular}} 
}
\startdata
x1y4z0.5r32      &  $0.63$ & $0.43$  & $0.2$  & $200$ & $150$ & $6.4^{+3.6}_{    }$ & $0.3$ & $0.0098$  &
$0.0067$ & $0.0006$ & $0.0005$ \\
x1y4z0.5r64      &  $0.43$ & $0.31$  & $0.12$  & $200$ & $150$ & $7.6^{+0.2}_{   }$ & $0.5$ & $0.0066$  &
$0.0046$ & $0.0005$ & $0.0002$ \\ 
x1y4z0.5r128     &  $0.19$ & $0.14$  & $0.05$  & $200$ & $150$ & $7.8^{   }_{-1.8}$ & $0.9^{+0.1}$ & $0.0029$  &
$0.0021$ & $0.0002$ & $0.0001$ \\ 
\hline
x0.5y2z0.5r32  &  $0.44$ & $0.35$  & $0.09$  & $300$ & $200$ & $6.2^{+3.8}_{    }$  & $0.3$ & $0.0084$ &
$0.0035$ & $0.0014$ & $0.0013$ \\
x0.5y2z0.5r64  &  $0.32$ & $0.25$  & $0.07$  & $300$ & $200$ & $11.0^{    }_{-6.0}$ & $0.5_{-0.1}    $ & $0.0055$  &
$0.0025$ & $0.0004$ & $0.0004$ \\
x0.5y2z0.5r128 &  $0.19$ & $0.15$  & $0.04$  & $300$ & $200$ & $9.6^{+1.4}_{    }$  & $0.9^{+0.1}    $ & $0.0032$  &
$0.0015$ & $0.0001$ & $0.0001$ \\
\hdashline
x1y4z1r32      &  $1.12$ & $0.85$  & $0.27$  & $300$ & $200$ & $5.2^{+4.8}_{    }$  & $0.5$ & $0.0190$  &
$0.0085$ & $0.0017$ & $0.0016$ \\
x1y4z1r64      &  $0.54$ & $0.41$  & $0.13$  & $300$ & $200$ & $7.6^{+5.8}_{-0.2}$  &
$1.0^{+0.1}_{-0.1}$ & $0.0088$  &
$0.0041$ & $0.0004$ & $0.0004$ \\ 
x1y4z1r128     &  $0.30$ & $0.26$  & $0.07$  & $200$ & $100$ & $15.4^{    }_{-2.4}$ &
$2.1^{+0.1}_{-0.1}$ & $0.0050$  &
$0.0023$ & $0.0001$ & $0.0001$ \\ 
\hline
x0.5y2z1r32    &  $0.89$ & $0.73$  & $0.16$  & $300$ & $200$ & $6.6^{+0.4}_{-0.2}$  & $0.6^{+0.1}$ & $0.0176$  &
$0.0073$  & $0.0042$ & $0.0041$ \\ 
x0.5y2z1r64   &  $0.57$ & $0.47$  & $0.10$  & $300$ & $200$ & $8.4^{+0.2}_{-1.0}$   &
$0.9^{+0.1}_{-0.1}$ & $0.0104$  &
$0.0047$  & $0.0013$ & $0.0013$ \\ 
x0.5y2z1r128  &  $0.29$ & $0.24$  & $0.05$  & $250$ & $150$ & $12.6^{+4.2}_{    }$  &
$1.6^{+0.3}_{-0.2}$ & $0.0053$  &
$0.0024$  & $0.0004$ & $0.0004$ \\ 
\hdashline
x1y4z2r32     &  $1.81$ & $1.44$  & $0.37$  & $300$ & $200$ & $9.0^{+2.8}_{}    $  & $1.1^{+0.1}$ & $0.0333$  &
$0.0144$  & $0.0049$ & $0.0047$ \\ 
x1y4z2r64     &  $0.86$ & $0.68$  & $0.17$  & $300$ & $200$ & $12.6^{+4.0}_{-2.2}$ &
$1.8^{+0.2}_{-0.1}$ & $0.0152$  &
$0.0068$  & $0.0010$ & $0.0009$ \\ 
x1y4z2r128    &  $0.39$ & $0.31$  & $0.08$  & $300$ & $150$ & $13.0^{+6.0}_{-0.4}$ &
$1.6^{+0.1}_{-0.1}$ & $0.0068$  &
$0.0031$  & $0.0002$ & $0.0002$ \\ 
\hline
x1y4z2.5r32   &  $2.27$ & $1.81$  & $0.46$  & $300$ & $200$ & $10.6^{+2.8}_{-0.2}$ & $1.4^{+0.1}$ & $0.0422$  &
$0.0181$  & $0.0067$ & $0.0065$ \\ 
x1y4z2.5r64   &  $1.30$ & $1.05$  & $0.25$  & $250$ & $150$ & $12.6^{+0.8}_{-0.4}$ & $1.9^{+0.1}$ & $0.0239$  &
$0.0105$  & $0.0022$ & $0.0021$ \\ 
x1y4z2.5r128  &  $1.31$ & $1.08$  & $0.22$  & $200$ & $100$ & $20.0^{   }_{-3.4}$  &
$2.0^{+0.1}_{-0.5}$ & $0.0202$  &
$0.0087$  & $0.0020$ & $0.0020$ \\ 
\hline
x1y4z3r32     &  $3.11$ & $2.51$  & $0.60$  & $300$ & $200$ & $13.4^{+2.0}_{-1.6}$ & $1.6^{+0.2}$ & $0.0618$  &
$0.0251$  & $0.0146$ & $0.0141$ \\ 
x1y4z3r64     &  $6.09$ & $5.08$  & $1.01$  & $270$ & $170$ & $27.4^{   }_{-2.4}$  &
$2.0^{+0.3}_{-0.3}$ & $0.1585$  &
$0.0508$  & $0.0628$ & $0.0614$ \\ 
x1y4z3r128    &  $9.13$ & $7.79$  & $1.34$  & $150$ & $100$ & $33.2^{   }_{    }$  &
$2.5^{+0.3}_{-0.4}$ & $0.2496$  &
$0.0739$  & $0.1048$ & $0.1028$ \\ 
\hline
x1y4z3.5r32   &  $3.50$ & $2.83$  & $0.68$  & $300$ & $200$ & $25.0^{+6.1}_{-4.4}$ &
$1.9^{+0.1}_{-0.1}$ & $0.0753$  &
$0.0281$  & $0.0234$ & $0.0228$ \\ 
x1y4z3.5r64   &  $6.35$ & $5.29$  & $1.05$  & $300$ & $200$ & $27.8^{+5.6}_{-2.8}$ &
$2.7^{+0.3}_{-0.3}$ & $0.1630$  &
$0.0529$  & $0.0626$ & $0.0612$ \\
\hline
x0.25y1z1r64  &  $0.42$ & $0.35$  & $0.07$  & $200$ & $100$ & $19.6^{+7.0}_{}$     &
$2.9^{+0.2}_{-0.3}$ & $0.0118$  &
$0.0035$ & $0.0061$ & $0.0060$ \\
x0.25y1z1r128 &  $0.49$ & $0.41$  & $0.08$  & $300$ & $200$ & $19.0^{+3.2}_{-6.4}$ &
$2.6^{+0.1}_{-0.2}$  & $0.0125$  &
$0.0041$ & $0.0052$ & $0.0051$ \\
x0.25y1z1r256 &  $0.51$ & $0.43$  & $0.08$  & $300$ & $200$ & $27.7^{   }_{-5.0}$  &
$3.4^{+0.6}_{-0.7}$  & $0.0161$  &
$0.0051$ & $0.0065$ & $0.0064$ \\ 
\hdashline
x0.5y2z2r32   &  $1.44$ & $1.17$  & $0.27$  & $300$ & $200$ & $17.4^{+4.6}_{-0.8}$ &
$2.4^{+0.2}_{-0.2}$ & $0.0370$  &
$0.0117$ & $0.0177$ & $0.0173$ \\
x0.5y2z2r64   &  $1.83$ & $1.51$  & $0.32$  & $300$ & $200$ & $25.0^{+8.4}_{   }$  &
$3.2^{+0.2}_{-0.1}$ & $0.0489$  &
$0.0151$ & $0.0224$ & $0.0219$ \\
x0.5y2z2r128  &  $2.13$ & $1.80$  & $0.33$  & $160$ & $100$ & $25.0^{+5.0}_{   }$  &
$3.1^{+0.6}_{-0.3}$ & $0.0633$  &
$0.0183$ & $0.0289$ & $0.0284$ \\
\hdashline
x1y4z4r32     &  $5.16$ & $4.17$  & $0.99$  & $300$ & $200$ & $22.8^{+10.6}_{    }$ &
$3.0^{+0.2}_{-0.5}$ & $0.1301$  &
$0.0417$ & $0.0565$ & $0.0553$ \\
x1y4z4r64     &  $7.85$ & $6.49$  & $1.36$  & $300$ & $200$ & $30.8^{+2.6}_{-8.6}$  &
$3.6^{+0.3}_{-0.3}$ & $0.2562$  &
$0.0649$ & $0.1374$ & $0.1353$ \\
x1y4z4r128    &  $8.41$ & $7.14$  & $1.26$  & $200$ & $100$ & $30.0^{~}_{~}$        &
$3.3^{+0.5}_{-0.7}$ & $0.2730$  &
$0.0714$ & $0.1331$ & $0.1310$ \\
\hdashline
x1y4z4r128pm8  & $9.25$  & $7.90$  & $1.35$  & $120$ & $100$ & $19.4^{   }_{  }$    &
$4.4^{0.7}_{-1.1}$ & $0.2716$  &
$0.0590$ & $0.0767$ & $0.0752$ \\
x1y4z4r32pm4  & $4.90$  & $3.99$  & $0.91$  & $300$ & $200$ & $22.8^{+5.8}_{-0.6}$  &
$2.6^{+0.6}_{-0.7}$ & $0.1188$  &
$0.0399$ & $0.0387$ & $0.0378$ \\
x1y4z4r64pm4  & $7.11$  & $5.90$  & $1.21$  & $300$ & $200$ & $30.0^{   }_{-10.0 }$ &
$3.7^{+0.2}_{-0.6}$ & $0.2053$  &
$0.0597$ & $0.0989$ & $0.0972$ \\
x1y4z4r128pm4  & $8.53$  & $7.24$  & $1.29$  & $200$ & $100$ & $22.6^{+4.8}_{  }$   &
$3.8^{+0.3}_{-0.3}$ & $0.2307$  &
$0.0604$ & $0.0942$ & $0.0924$ \\
x1y4z4r128pm2  & $5.39$ & $4.51$  & $0.88$  & $300$ & $200$ & $32.8^{    }_{    }$  &
$3.6^{+0.5}_{-1.1}$ & $0.1421$  &
$0.0439$ & $0.0464$ & $0.0455$ \\
x1y4z4r128pm1  & $ - $  & $ - $   & $ - $   & $300$ & $ - $ & $ - $ & $ - $  & $ - $ & $ - $ & $ - $ & $ - $ \\
\hdashline
x2y8z8r32     &  $20.06$ & $16.08$ & $3.98$  & $300$ & $200$ & $34.4^{~}_{-2.4}$      &
$3.9^{+0.6}_{-0.6}$ & $0.6387$  &
$0.1701$ & $0.3401$ & $0.3351$ \\
x2y8z8r64     &  $21.52$ & $17.53$ & $3.99$  & $200$ & $150$ & $47.0^{+19.6}_{-11.0}$ &
$3.9^{+0.5}_{-0.7}$ & $0.6823$  &
$0.1791$ & $0.3451$ & $0.3401$ \\
\hline
x1y4z6r32        &  $5.93$ & $4.79$  & $1.14$  & $300$ & $250$ & $47.6^{+2.4}_{-7.4}$ &
$4.8^{+0.3}_{-0.3}$ & $0.1558$ &
$0.0477$ & $0.0725$ & $0.0710$ \\
x1y4z6r64        &  $7.60$ & $6.30$  & $1.31$  & $280$ & $180$ & $37.0^{+3.0}_{-2.2}$ &
$6.9^{+0.5}_{-0.4}$ & $0.2390$ &
$0.0632$ & $0.1228$ & $0.1207$ \\
\hline
x0.25y1z2r64     &  $0.42$ & $0.34$  & $0.08$  & $300$ & $200$ & $33.4^{+6.6}_{    }$ &
$7.0^{+0.4}_{-0.5}$ & $0.0146$ &
$0.0034$ & $0.0091$ & $0.0090$  \\
x0.25y1z2r128    &  $0.48$ & $0.40$  & $0.08$  & $250$ & $150$ & $50.0^{+3.4}_{-1.4}$ &
$9.2^{+1.4}_{-1.9}$ & $0.0211$ &
$0.0040$ & $0.0151$ & $0.0150$  \\
x0.25y1z2r256    &  $0.52$ & $0.44$  & $0.08$  & $200$ & $100$ & $59.6^{+19.2}_{    }$&
$9.6^{+2.2}_{-1.1}$ & $0.0531$ &
$0.0044$ & $0.0143$ & $0.0141$  \\
\hdashline
x0.5y2z4r32      &  $1.53$ & $1.24$  & $0.28$  & $300$ & $200$ & $40.0^{    }_{-11.4}$ &
$6.0^{+0.2}_{-0.1}$ & $0.0451$ &
$0.0124$ & $0.0250$ & $0.0245$ \\
x0.5y2z4r64      &  $2.04$ & $1.68$  & $0.36$  & $300$ & $200$ & $44.3^{    }_{-11.0}$ &
$8.8^{+1.4}_{-0.4}$ & $0.0648$ &
$0.0168$ & $0.0359$ & $0.0353$ \\
x0.5y2z4r128     &  $2.13$ & $1.79$  & $0.35$  & $200$ & $150$ & $61.4^{+5.0}_{-11.4}$ &
$12.0^{0.4}_{-1.8}$ & $0.0918$ &
$0.0179$ & $0.0590$ & $0.0583$ \\
\hdashline
x1y4z8r32        &  $6.10$ & $4.88$  & $1.22$  & $300$ & $200$ & $55.0^{+2.2}_{-7.2}$ &
$9.7^{+0.8}_{-2.3}$ & $0.2261$ &
$0.0488$ & $0.1428$ & $0.1409$ \\ 
x1y4z8r32by      &  $5.97$ & $4.78$  & $1.19$  & $300$ & $200$ & $50.0^{+6.2}_{    }$ &
$8.3^{+0.7}_{-1.1}$ & $0.2048$ &
$0.0478$ & $0.1225$ & $0.1207$ \\ 
x1y4z8r64        &  $7.21$ & $5.92$  & $1.29$  & $200$ & $100$ & $50.0^{+3.4}_{    }$ &
$7.6^{+1.1}_{-2.5}$ & $0.3833$ &
$0.0595$ & $0.1850$ & $0.1827$ \\ 
x1y4z8r128       &  $7.92$ & $6.66$  & $1.26$  & $200$ & $100$ & $62.0^{+20.0}_{   }$ &
$9.2^{+3.0}_{-1.3}$ & $0.3790$ &
$0.0680$ & $0.2491$ & $0.2470$ \\ 
\hline 
x1y4z10r32    &  $5.68$ & $4.54$  & $1.13$  & $300$ & $200$ & $57.2^{+9.6}_{-7.2}$ &
$11.3^{+0.5}_{-0.4}$ & $0.1949$ &
$0.0454$  & $0.1178$ & $0.1160$ \\
x1y4z10r64    &  $7.16$ & $5.91$  & $1.25$  & $300$ & $200$ & $80.0^{    }_{-7.0}$ &
$12.5^{+1.6}_{-2.0}$ & $0.2444$ &
$0.0591$  & $0.1357$ & $0.1338$ \\ 
\hline
x1y4z12r32    &  $6.29$ & $5.04$  & $1.25$  & $300$ & $200$ & $96.8^{+4.0}_{    }$ &
$15.7^{+2.8}_{-1.0}$ & $0.2000$ &
$0.0504$  & $0.1142$ & $0.1122$ \\
x1y4z12r64    &  $7.00$ & $5.73$  & $1.27$  & $300$ & $200$ & $100.0^{    }_{-5.0}$&
$16.9^{+1.1}_{-3.1}$ & $0.3349$ &
$0.0574$  & $0.2306$ & $0.2285$  
\enddata
\tablenotetext{(a)}{~Name convention: x{\it{n}}~y{\it{m}}~z{\it{k}} denotes box dimension
($L_x$,$L_y$,$L_z$)$=$({\it{n}},{\it{m}},{\it{k}})$H$; `r$32$' means typical resolution $32/H$,
`r$64$' and
`r$128$' are $64/H$ and $128/H$ respectively; `pm8', `pm4',`pm2', and `pm1' denote resistive runs with
magnetic Prandtl number $\rm{Pm} = 8$,$4$,$2$, and $1$ for fixed Reynolds number $\rm{Re} = 3125$;  
`by' stands for an initial azimuthal magnetic field $\mathbf{B}(t=0)=B_0
\sin(2\pi x/\lx)\hat{\mathbf{y}}$ instead of the vertical configuration
$\mathbf{B}(t=0)=B_0\sin(2\pi x/\lx)\hat{\mathbf{z}}$ used in most runs.}
\tablenotetext{(b)}{~$\alpha_{\rm{M}}$ and $\alpha_{\rm R}$ are the Maxwell and Reynolds stresses 
normalized by $\rho_0 c_s^2$; $\alpha_{\rm tot}$ is the sum of these two.} 
\tablenotetext{(c)}{~Duration of the simulation.} 
\tablenotetext{(d)}{~Last $\Delta T_{\rm avg}$ orbits chosen for time average.}
\tablenotetext{(e)}{~The period of the oscillating magnetic field cycle. Measured based on the power
spectra of the largest vertical mode ($k=2\pi/L_z$) of $\ob{B_y}(t)$. We separate the time
sequence to two parts, and the errors are derived from the range of $T_{\rm cycle}$ over two different
parts.}
\tablenotetext{(f)}{~The time lag between the azimuthal and radial mean field defined in
section~\ref{sec:cycle} and used in Equation~\ref{eq:phase_lag} for calculating the phase lag.}
\tablenotetext{(g)}{~The total magnetic energy $\langle\langle B^2/8\pi P_0\rangle\rangle_t$, total
kinetic energy $\langle\langle \rho v^2/2 P_0\rangle\rangle_t$, total mean field energy
$\langle\langle\ob{B}^2/8\pi P_0\rangle\rangle_t$, and azimuthal mean field energy
$\langle\langle\ob{B}_y^2/8\pi P_0\rangle\rangle_t$.}
\end{deluxetable}}

\clearpage
{\renewcommand{\arraystretch}{1.4}
\begin{deluxetable}{lccc}
\tabletypesize{\footnotesize}
\tablecolumns{4} \tablewidth{0pc}
\tablecaption{Time and volume averaged magnetic field related quantities \label{tab:tab2}}
\tablehead{\colhead{ } & \colhead{x1y4z1r128} & \colhead{x1y4z4r128} & \colhead{x1y4z8r128} }
\startdata
$\langle\langle   B^2/8\pi P_0\rangle\rangle_t$         &  $0.0050$  & $0.2730$ & $0.3790$\\
$\langle\langle B_x^2/8\pi P_0\rangle\rangle_t$         &  $0.0007$  & $0.0352$ & $0.0357$\\
$\langle\langle B_y^2/8\pi P_0\rangle\rangle_t$         &  $0.0041$  & $0.2208$ & $0.3273$\\
$\langle\langle B_z^2/8\pi P_0\rangle\rangle_t$         &  $0.0003$  & $0.0170$ & $0.0160$\\
$\langle\langle \ob{B}^2  /8\pi P_0\rangle\rangle_t$    &  $0.0001$  & $0.1331$ & $0.2491$ \\
$\langle\langle \ob{B}_x^2/8\pi P_0\rangle\rangle_t$    &  $0.0$     & $0.0020$ & $0.0022$ \\
$\langle\langle \ob{B}_y^2/8\pi P_0\rangle\rangle_t$    &  $0.0001$  & $0.1311$ & $0.2470$ \\
$\langle\langle \ob{B}_z^2/8\pi P_0\rangle\rangle_t$    &  $0.0$     & $0.0  $  & $0.0$    \\
$\langle\langle {b}^2  /8\pi P_0\rangle\rangle_t$       &  $0.0049$  & $0.1399$ & $0.1299$ \\
$\langle\langle {b}_x^2/8\pi P_0\rangle\rangle_t$       &  $0.0007$  & $0.0332$ & $0.0335$ \\
$\langle\langle {b}_y^2/8\pi P_0\rangle\rangle_t$       &  $0.0040$  & $0.0897$ & $0.0803$ \\
$\langle\langle {b}_z^2/8\pi P_0\rangle\rangle_t$       &  $0.0003$  & $0.0170$ & $0.0160$ \\
$\langle\langle - B_x B_y/4\pi P_0\rangle\rangle_t$           &  $0.0023$  & $0.0714$ & $0.0680$ \\
$\langle\langle - \ob{B}_x\ob{B}_y/4\pi P_0\rangle\rangle_t$  &  $0.0$     & $0.0115$ & $0.0110$ \\
$\langle\langle - b_x b_y/4\pi P_0\rangle\rangle_t$           &  $0.0023$  & $0.0599$ & $0.0570$ \\
\hline
$\langle\langle \rho v^2/2P_0 \rangle\rangle_t$               &  $0.0028$  & $0.0707$ & $0.0706$ \\
$\langle\langle \rho v_x^2/2P_0 \rangle\rangle_t$             &  $0.0014$  & $0.0226$ & $0.0226$ \\
$\langle\langle \rho \delta v_y^2/2P_0 \rangle\rangle_t$      &  $0.0009$  & $0.0365$ & $0.0369$ \\
$\langle\langle \rho v_z^2/2P_0 \rangle\rangle_t$             &  $0.0005$  & $0.0116$ & $0.0111$ \\
$\langle\langle \rho v_x\delta v_y/P_0 \rangle\rangle_t$      &  $0.0007$  & $0.0126$ & $0.0131$ \\
$\langle\langle (\delta\rho /\rho_0)^2 \rangle^{1/2}\rangle_t$    &  $0.0312$  & $0.1990$ & $0.2243$ \\

\enddata
\end{deluxetable}
}
\clearpage

\bibliographystyle{mnras}

\begin{thebibliography}{}
\makeatletter
\relax
\def\mn@urlcharsother{\let\do\@makeother \do\$\do\&\do\#\do\^\do\_\do\%\do\~}
\def\mn@doi{\begingroup\mn@urlcharsother \@ifnextchar [ {\mn@doi@}
  {\mn@doi@[]}}
\def\mn@doi@[#1]#2{\def\@tempa{#1}\ifx\@tempa\@empty \href
  {http://dx.doi.org/#2} {doi:#2}\else \href {http://dx.doi.org/#2} {#1}\fi
  \endgroup}
\def\mn@eprint#1#2{\mn@eprint@#1:#2::\@nil}
\def\mn@eprint@arXiv#1{\href {http://arxiv.org/abs/#1} {{\tt arXiv:#1}}}
\def\mn@eprint@dblp#1{\href {http://dblp.uni-trier.de/rec/bibtex/#1.xml}
  {dblp:#1}}
\def\mn@eprint@#1:#2:#3:#4\@nil{\def\@tempa {#1}\def\@tempb {#2}\def\@tempc
  {#3}\ifx \@tempc \@empty \let \@tempc \@tempb \let \@tempb \@tempa \fi \ifx
  \@tempb \@empty \def\@tempb {arXiv}\fi \@ifundefined
  {mn@eprint@\@tempb}{\@tempb:\@tempc}{\expandafter \expandafter \csname
  mn@eprint@\@tempb\endcsname \expandafter{\@tempc}}}

\bibitem[\protect\citeauthoryear{{Abramowicz}, {Brandenburg}  \&
  {Lasota}}{{Abramowicz} et~al.}{1996}]{abl1996}
{Abramowicz} M.,  {Brandenburg} A.,   {Lasota} J.-P.,  1996, \mnras, \href
  {http://adsabs.harvard.edu/abs/1996MNRAS.281L..21A} {281, L21}

\bibitem[\protect\citeauthoryear{{Bai}}{{Bai}}{2014}]{Bai2014}
{Bai} X.-N.,  2014, \mn@doi [\apj] {10.1088/0004-637X/791/2/137}, \href
  {http://adsabs.harvard.edu/abs/2014ApJ...791..137B} {791, 137}

\bibitem[\protect\citeauthoryear{{Bai}}{{Bai}}{2015}]{Bai2015}
{Bai} X.-N.,  2015, \mn@doi [\apj] {10.1088/0004-637X/798/2/84}, \href
  {http://adsabs.harvard.edu/abs/2015ApJ...798...84B} {798, 84}

\bibitem[\protect\citeauthoryear{{Bai} \& {Stone}}{{Bai} \&
  {Stone}}{2011}]{BS2011}
{Bai} X.-N.,  {Stone} J.~M.,  2011, \mn@doi [\apj]
  {10.1088/0004-637X/736/2/144}, \href
  {http://adsabs.harvard.edu/abs/2011ApJ...736..144B} {736, 144}

\bibitem[\protect\citeauthoryear{{Bai} \& {Stone}}{{Bai} \&
  {Stone}}{2013a}]{bai_stone2013}
{Bai} X.-N.,  {Stone} J.~M.,  2013a, \mn@doi [\apj]
  {10.1088/0004-637X/767/1/30}, \href
  {http://adsabs.harvard.edu/abs/2013ApJ...767...30B} {767, 30}

\bibitem[\protect\citeauthoryear{{Bai} \& {Stone}}{{Bai} \&
  {Stone}}{2013b}]{BS2013}
{Bai} X.-N.,  {Stone} J.~M.,  2013b, \mn@doi [\apj]
  {10.1088/0004-637X/769/1/76}, \href
  {http://adsabs.harvard.edu/abs/2013ApJ...769...76B} {769, 76}

\bibitem[\protect\citeauthoryear{{Blackman} \& {Nauman}}{{Blackman} \&
  {Nauman}}{2015}]{BN2015}
{Blackman} E.~G.,  {Nauman} F.,  2015, preprint, \href
  {http://adsabs.harvard.edu/abs/2015arXiv150100291B} {} (\mn@eprint {arXiv}
  {1501.00291})

\bibitem[\protect\citeauthoryear{{Blackman} \& {Tan}}{{Blackman} \&
  {Tan}}{2004}]{BT2004}
{Blackman} E.~G.,  {Tan} J.~C.,  2004, \mn@doi [\apss]
  {10.1023/B:ASTR.0000045043.87692.4a}, \href
  {http://adsabs.harvard.edu/abs/2004Ap%26SS.292..395B} {292, 395}


\bibitem[\protect\citeauthoryear{{Blackman}, {Penna}  \&
  {Varni{\`e}re}}{{Blackman} et~al.}{2008}]{blackman2008}
{Blackman} E.~G.,  {Penna} R.~F.,   {Varni{\`e}re} P.,  2008, \mn@doi [\na]
  {10.1016/j.newast.2007.10.004}, \href
  {http://adsabs.harvard.edu/abs/2008NewA...13..244B} {13, 244}

\bibitem[\protect\citeauthoryear{{Bodo}, {Cattaneo}, {Ferrari}, {Mignone}  \&
  {Rossi}}{{Bodo} et~al.}{2011}]{bodoetal2011}
{Bodo} G.,  {Cattaneo} F.,  {Ferrari} A.,  {Mignone} A.,   {Rossi} P.,  2011,
  \mn@doi [\apj] {10.1088/0004-637X/739/2/82}, \href
  {http://adsabs.harvard.edu/abs/2011ApJ...739...82B} {739, 82}

\bibitem[\protect\citeauthoryear{{Bodo}, {Cattaneo}, {Mignone}  \&
  {Rossi}}{{Bodo} et~al.}{2014}]{Bodoetal2014}
{Bodo} G.,  {Cattaneo} F.,  {Mignone} A.,   {Rossi} P.,  2014, \mn@doi [\apjl]
  {10.1088/2041-8205/787/1/L13}, \href
  {http://adsabs.harvard.edu/abs/2014ApJ...787L..13B} {787, L13}

\bibitem[\protect\citeauthoryear{{Brandenburg} \& {Subramanian}}{{Brandenburg}
  \& {Subramanian}}{2005}]{BS2005}
{Brandenburg} A.,  {Subramanian} K.,  2005, \mn@doi [\physrep]
  {10.1016/j.physrep.2005.06.005}, \href
  {http://adsabs.harvard.edu/abs/2005PhR...417....1B} {417, 1}

\bibitem[\protect\citeauthoryear{{Brandenburg}, {Nordlund}, {Stein}  \&
  {Torkelsson}}{{Brandenburg} et~al.}{1995}]{Brandenburgetal1995}
{Brandenburg} A.,  {Nordlund} A.,  {Stein} R.~F.,   {Torkelsson} U.,  1995,
  \mn@doi [\apj] {10.1086/175831}, \href
  {http://adsabs.harvard.edu/abs/1995ApJ...446..741B} {446, 741}

\bibitem[\protect\citeauthoryear{{Davis}, {Stone}  \& {Pessah}}{{Davis}
  et~al.}{2010}]{davisetal2010}
{Davis} S.~W.,  {Stone} J.~M.,   {Pessah} M.~E.,  2010, \mn@doi [\apj]
  {10.1088/0004-637X/713/1/52}, \href
  {http://adsabs.harvard.edu/abs/2010ApJ...713...52D} {713, 52}

\bibitem[\protect\citeauthoryear{{Fromang}}{{Fromang}}{2010}]{Fromang2010}
{Fromang} S.,  2010, \mn@doi [\aap] {10.1051/0004-6361/201014284}, \href
  {http://adsabs.harvard.edu/abs/2010A%26A...514L...5F} {514, L5}


\bibitem[\protect\citeauthoryear{{Fromang} \& {Papaloizou}}{{Fromang} \&
  {Papaloizou}}{2007}]{FP2007}
{Fromang} S.,  {Papaloizou} J.,  2007, \mn@doi [\aap]
  {10.1051/0004-6361:20077942}, \href
  {http://adsabs.harvard.edu/abs/2007A%26A...476.1113F} {476, 1113}


\bibitem[\protect\citeauthoryear{{Fromang}, {Papaloizou}, {Lesur}  \&
  {Heinemann}}{{Fromang} et~al.}{2007}]{FPLH2007}
{Fromang} S.,  {Papaloizou} J.,  {Lesur} G.,   {Heinemann} T.,  2007, \mn@doi
  [\aap] {10.1051/0004-6361:20077943}, \href
  {http://adsabs.harvard.edu/abs/2007A%26A...476.1123F} {476, 1123}


\bibitem[\protect\citeauthoryear{{Fromang}, {Latter}, {Lesur}  \&
  {Ogilvie}}{{Fromang} et~al.}{2013}]{FLLO2013}
{Fromang} S.,  {Latter} H.,  {Lesur} G.,   {Ogilvie} G.~I.,  2013, \mn@doi
  [\aap] {10.1051/0004-6361/201220016}, \href
  {http://adsabs.harvard.edu/abs/2013A%26A...552A..71F} {552, A71}


\bibitem[\protect\citeauthoryear{{Gressel}}{{Gressel}}{2010}]{Gressel2010}
{Gressel} O.,  2010, \mn@doi [\mnras] {10.1111/j.1365-2966.2010.16440.x}, \href
  {http://adsabs.harvard.edu/abs/2010MNRAS.405...41G} {405, 41}

\bibitem[\protect\citeauthoryear{{Gressel} \& {Pessah}}{{Gressel} \&
  {Pessah}}{2015}]{GP2015}
{Gressel} O.,  {Pessah} M.~E.,  2015, \mn@doi [\apj]
  {10.1088/0004-637X/810/1/59}, \href
  {http://adsabs.harvard.edu/abs/2015ApJ...810...59G} {810, 59}

\bibitem[\protect\citeauthoryear{{Gressel} \& {Ziegler}}{{Gressel} \&
  {Ziegler}}{2007}]{GZ2007}
{Gressel} O.,  {Ziegler} U.,  2007, \mn@doi [Computer Physics Communications]
  {10.1016/j.cpc.2007.01.010}, \href
  {http://adsabs.harvard.edu/abs/2007CoPhC.176..652G} {176, 652}

\bibitem[\protect\citeauthoryear{{Guan} \& {Gammie}}{{Guan} \&
  {Gammie}}{2011}]{GG2011}
{Guan} X.,  {Gammie} C.~F.,  2011, \mn@doi [\apj]
  {10.1088/0004-637X/728/2/130}, \href
  {http://adsabs.harvard.edu/abs/2011ApJ...728..130G} {728, 130}

\bibitem[\protect\citeauthoryear{{Guan}, {Gammie}, {Simon}  \&
  {Johnson}}{{Guan} et~al.}{2009}]{Guan2009}
{Guan} X.,  {Gammie} C.~F.,  {Simon} J.~B.,   {Johnson} B.~M.,  2009, \mn@doi
  [\apj] {10.1088/0004-637X/694/2/1010}, \href
  {http://adsabs.harvard.edu/abs/2009ApJ...694.1010G} {694, 1010}

\bibitem[\protect\citeauthoryear{{Hawley} \& {Stone}}{{Hawley} \&
  {Stone}}{1998}]{HS1998}
{Hawley} J.~F.,  {Stone} J.~M.,  1998, \mn@doi [\apj] {10.1086/305849}, \href
  {http://adsabs.harvard.edu/abs/1998ApJ...501..758H} {501, 758}

\bibitem[\protect\citeauthoryear{{Hawley}, {Gammie}  \& {Balbus}}{{Hawley}
  et~al.}{1995}]{hgb95}
{Hawley} J.~F.,  {Gammie} C.~F.,   {Balbus} S.~A.,  1995, \mn@doi [\apj]
  {10.1086/175311}, \href {http://adsabs.harvard.edu/abs/1995ApJ...440..742H}
  {440, 742}

\bibitem[\protect\citeauthoryear{{Hawley}, {Gammie}  \& {Balbus}}{{Hawley}
  et~al.}{1996}]{hgb96}
{Hawley} J.~F.,  {Gammie} C.~F.,   {Balbus} S.~A.,  1996, \mn@doi [\apj]
  {10.1086/177356}, \href {http://adsabs.harvard.edu/abs/1996ApJ...464..690H}
  {464, 690}

\bibitem[\protect\citeauthoryear{{Hawley}, {Balbus}  \& {Winters}}{{Hawley}
  et~al.}{1999}]{hbw1999}
{Hawley} J.~F.,  {Balbus} S.~A.,   {Winters} W.~F.,  1999, \mn@doi [\apj]
  {10.1086/307282}, \href {http://adsabs.harvard.edu/abs/1999ApJ...518..394H}
  {518, 394}

\bibitem[\protect\citeauthoryear{{Hawley}, {Guan}  \& {Krolik}}{{Hawley}
  et~al.}{2011}]{hgk2011}
{Hawley} J.~F.,  {Guan} X.,   {Krolik} J.~H.,  2011, \mn@doi [\apj]
  {10.1088/0004-637X/738/1/84}, \href
  {http://adsabs.harvard.edu/abs/2011ApJ...738...84H} {738, 84}

\bibitem[\protect\citeauthoryear{{Heinemann} \& {Papaloizou}}{{Heinemann} \&
  {Papaloizou}}{2009}]{HP2009}
{Heinemann} T.,  {Papaloizou} J.~C.~B.,  2009, \mn@doi [\mnras]
  {10.1111/j.1365-2966.2009.14800.x}, \href
  {http://adsabs.harvard.edu/abs/2009MNRAS.397...64H} {397, 64}

\bibitem[\protect\citeauthoryear{{Herault}, {Rincon}, {Cossu}, {Lesur},
  {Ogilvie}  \& {Longaretti}}{{Herault} et~al.}{2011}]{Heraultetal2011}
{Herault} J.,  {Rincon} F.,  {Cossu} C.,  {Lesur} G.,  {Ogilvie} G.~I.,
  {Longaretti} P.-Y.,  2011, \mn@doi [\pre] {10.1103/PhysRevE.84.036321}, \href
  {http://adsabs.harvard.edu/abs/2011PhRvE..84c6321H} {84, 036321}

\bibitem[\protect\citeauthoryear{{Johnson}, {Guan}  \& {Gammie}}{{Johnson}
  et~al.}{2008}]{Johnson2008}
{Johnson} B.~M.,  {Guan} X.,   {Gammie} C.~F.,  2008, \mn@doi [\apjs]
  {10.1086/586707}, \href {http://adsabs.harvard.edu/abs/2008ApJS..177..373J}
  {177, 373}

\bibitem[\protect\citeauthoryear{{K{\"a}pyl{\"a}}, {Mantere}, {Cole},
  {Warnecke}  \& {Brandenburg}}{{K{\"a}pyl{\"a}} et~al.}{2013}]{kapylaetal2013}
{K{\"a}pyl{\"a}} P.~J.,  {Mantere} M.~J.,  {Cole} E.,  {Warnecke} J.,
  {Brandenburg} A.,  2013, \mn@doi [\apj] {10.1088/0004-637X/778/1/41}, \href
  {http://adsabs.harvard.edu/abs/2013ApJ...778...41K} {778, 41}

\bibitem[\protect\citeauthoryear{{King}, {Pringle}  \& {Livio}}{{King}
  et~al.}{2007}]{kpl2007}
{King} A.~R.,  {Pringle} J.~E.,   {Livio} M.,  2007, \mn@doi [\mnras]
  {10.1111/j.1365-2966.2007.11556.x}, \href
  {http://adsabs.harvard.edu/abs/2007MNRAS.376.1740K} {376, 1740}

\bibitem[\protect\citeauthoryear{{Kunz} \& {Lesur}}{{Kunz} \&
  {Lesur}}{2013}]{kunz13}
{Kunz} M.~W.,  {Lesur} G.,  2013, \mn@doi [\mnras] {10.1093/mnras/stt1171},
  \href {http://adsabs.harvard.edu/abs/2013MNRAS.434.2295K} {434, 2295}

\bibitem[\protect\citeauthoryear{{Lesur} \& {Longaretti}}{{Lesur} \&
  {Longaretti}}{2007}]{LL2007}
{Lesur} G.,  {Longaretti} P.-Y.,  2007, \mn@doi [\mnras]
  {10.1111/j.1365-2966.2007.11888.x}, \href
  {http://adsabs.harvard.edu/abs/2007MNRAS.378.1471L} {378, 1471}

\bibitem[\protect\citeauthoryear{{Lesur} \& {Ogilvie}}{{Lesur} \&
  {Ogilvie}}{2008a}]{LO08-2}
{Lesur} G.,  {Ogilvie} G.~I.,  2008a, \mn@doi [\mnras]
  {10.1111/j.1365-2966.2008.13993.x}, \href
  {http://adsabs.harvard.edu/abs/2008MNRAS.391.1437L} {391, 1437}

\bibitem[\protect\citeauthoryear{{Lesur} \& {Ogilvie}}{{Lesur} \&
  {Ogilvie}}{2008b}]{LO08}
{Lesur} G.,  {Ogilvie} G.~I.,  2008b, \mn@doi [\aap]
  {10.1051/0004-6361:200810152}, \href
  {http://adsabs.harvard.edu/abs/2008A%26A...488..451L} {488, 451}


\bibitem[\protect\citeauthoryear{{Lesur}, {Ferreira}  \& {Ogilvie}}{{Lesur}
  et~al.}{2013}]{LFO2013}
{Lesur} G.,  {Ferreira} J.,   {Ogilvie} G.~I.,  2013, \mn@doi [\aap]
  {10.1051/0004-6361/201220395}, \href
  {http://adsabs.harvard.edu/abs/2013A%26A...550A..61L} {550, A61}


\bibitem[\protect\citeauthoryear{{Lesur}, {Kunz}  \& {Fromang}}{{Lesur}
  et~al.}{2014}]{lesur14}
{Lesur} G.,  {Kunz} M.~W.,   {Fromang} S.,  2014, \mn@doi [\aap]
  {10.1051/0004-6361/201423660}, \href
  {http://adsabs.harvard.edu/abs/2014A%26A...566A..56L} {566, A56}


\bibitem[\protect\citeauthoryear{{Masada} \& {Sano}}{{Masada} \&
  {Sano}}{2008}]{MS2008}
{Masada} Y.,  {Sano} T.,  2008, \mn@doi [\apj] {10.1086/592601}, \href
  {http://adsabs.harvard.edu/abs/2008ApJ...689.1234M} {689, 1234}

\bibitem[\protect\citeauthoryear{{Masset}}{{Masset}}{2000}]{Masset2000}
{Masset} F.,  2000, \mn@doi [\aaps] {10.1051/aas:2000116}, \href
  {http://adsabs.harvard.edu/abs/2000A%26AS..141..165M} {141, 165}


\bibitem[\protect\citeauthoryear{{Meheut}, {Fromang}, {Lesur}, {Joos}  \&
  {Longaretti}}{{Meheut} et~al.}{2015}]{Meheutetal2015}
{Meheut} H.,  {Fromang} S.,  {Lesur} G.,  {Joos} M.,   {Longaretti} P.-Y.,
  2015, \mn@doi [\aap] {10.1051/0004-6361/201525688}, \href
  {http://adsabs.harvard.edu/abs/2015A%26A...579A.117M} {579, A117}


\bibitem[\protect\citeauthoryear{{Miller} \& {Stone}}{{Miller} \&
  {Stone}}{2000}]{MS2000}
{Miller} K.~A.,  {Stone} J.~M.,  2000, \mn@doi [\apj] {10.1086/308736}, \href
  {http://adsabs.harvard.edu/abs/2000ApJ...534..398M} {534, 398}

\bibitem[\protect\citeauthoryear{{Oishi} \& {Mac Low}}{{Oishi} \& {Mac
  Low}}{2011}]{OM2011}
{Oishi} J.~S.,  {Mac Low} M.-M.,  2011, \mn@doi [\apj]
  {10.1088/0004-637X/740/1/18}, \href
  {http://adsabs.harvard.edu/abs/2011ApJ...740...18O} {740, 18}

\bibitem[\protect\citeauthoryear{{Pessah}, {Chan}  \& {Psaltis}}{{Pessah}
  et~al.}{2006}]{pessahetal2006}
{Pessah} M.~E.,  {Chan} C.-K.,   {Psaltis} D.,  2006, \mn@doi [\mnras]
  {10.1111/j.1365-2966.2006.10824.x}, \href
  {http://adsabs.harvard.edu/abs/2006MNRAS.372..183P} {372, 183}

\bibitem[\protect\citeauthoryear{{Pessah}, {Chan}  \& {Psaltis}}{{Pessah}
  et~al.}{2007}]{pessahetal2007}
{Pessah} M.~E.,  {Chan} C.-k.,   {Psaltis} D.,  2007, \mn@doi [\apjl]
  {10.1086/522585}, \href {http://adsabs.harvard.edu/abs/2007ApJ...668L..51P}
  {668, L51}

\bibitem[\protect\citeauthoryear{{Rempel}, {Lesur}  \& {Proctor}}{{Rempel}
  et~al.}{2010}]{Rempeletal2010}
{Rempel} E.~L.,  {Lesur} G.,   {Proctor} M.~R.~E.,  2010, \mn@doi [Physical
  Review Letters] {10.1103/PhysRevLett.105.044501}, \href
  {http://adsabs.harvard.edu/abs/2010PhRvL.105d4501R} {105, 044501}

\bibitem[\protect\citeauthoryear{{Rincon}, {Ogilvie}  \& {Proctor}}{{Rincon}
  et~al.}{2007}]{Rincon2007}
{Rincon} F.,  {Ogilvie} G.~I.,   {Proctor} M.~R.~E.,  2007, \mn@doi [Physical
  Review Letters] {10.1103/PhysRevLett.98.254502}, \href
  {http://adsabs.harvard.edu/abs/2007PhRvL..98y4502R} {98, 254502}

\bibitem[\protect\citeauthoryear{{Rincon}, {Ogilvie}, {Proctor}  \&
  {Cossu}}{{Rincon} et~al.}{2008}]{Rincon2008}
{Rincon} F.,  {Ogilvie} G.~I.,  {Proctor} M.~R.~E.,   {Cossu} C.,  2008,
  \mn@doi [Astronomische Nachrichten] {10.1002/asna.200811010}, \href
  {http://adsabs.harvard.edu/abs/2008AN....329..750R} {329, 750}

\bibitem[\protect\citeauthoryear{{Riols}, {Rincon}, {Cossu}, {Lesur},
  {Longaretti}, {Ogilvie}  \& {Herault}}{{Riols} et~al.}{2013}]{Riolsetal2013}
{Riols} A.,  {Rincon} F.,  {Cossu} C.,  {Lesur} G.,  {Longaretti} P.-Y.,
  {Ogilvie} G.~I.,   {Herault} J.,  2013, \mn@doi [Journal of Fluid Mechanics]
  {10.1017/jfm.2013.317}, \href
  {http://adsabs.harvard.edu/abs/2013JFM...731....1R} {731, 1}

\bibitem[\protect\citeauthoryear{{Riols}, {Rincon}, {Cossu}, {Lesur}, {Ogilvie}
   \& {Longaretti}}{{Riols} et~al.}{2015}]{Riolsetal2015}
{Riols} A.,  {Rincon} F.,  {Cossu} C.,  {Lesur} G.,  {Ogilvie} G.~I.,
  {Longaretti} P.-Y.,  2015, \mn@doi [\aap] {10.1051/0004-6361/201424324},
  \href {http://adsabs.harvard.edu/abs/2015A%26A...575A..14R} {575, A14}


\bibitem[\protect\citeauthoryear{{Rogachevskii} \& {Kleeorin}}{{Rogachevskii}
  \& {Kleeorin}}{2004}]{RogaKlee2004}
{Rogachevskii} I.,  {Kleeorin} N.,  2004, \mn@doi [\pre]
  {10.1103/PhysRevE.70.046310}, \href
  {http://adsabs.harvard.edu/abs/2004PhRvE..70d6310R} {70, 046310}

\bibitem[\protect\citeauthoryear{{R{\"u}diger} \& {Pipin}}{{R{\"u}diger} \&
  {Pipin}}{2000}]{RP2000}
{R{\"u}diger} G.,  {Pipin} V.~V.,  2000, \aap, \href
  {http://adsabs.harvard.edu/abs/2000A%26A...362..756R} {362, 756}


\bibitem[\protect\citeauthoryear{{Salmeron} \& {Wardle}}{{Salmeron} \&
  {Wardle}}{2003}]{SW2003}
{Salmeron} R.,  {Wardle} M.,  2003, \mn@doi [\mnras]
  {10.1046/j.1365-8711.2003.07024.x}, \href
  {http://adsabs.harvard.edu/abs/2003MNRAS.345..992S} {345, 992}

\bibitem[\protect\citeauthoryear{{Sano} \& {Stone}}{{Sano} \&
  {Stone}}{2002}]{sano2002}
{Sano} T.,  {Stone} J.~M.,  2002, \mn@doi [\apj] {10.1086/342172}, \href
  {http://adsabs.harvard.edu/abs/2002ApJ...577..534S} {577, 534}

\bibitem[\protect\citeauthoryear{{Sano}, {Inutsuka}, {Turner}  \&
  {Stone}}{{Sano} et~al.}{2004}]{sano2004}
{Sano} T.,  {Inutsuka} S.-i.,  {Turner} N.~J.,   {Stone} J.~M.,  2004, \mn@doi
  [\apj] {10.1086/382184}, \href
  {http://adsabs.harvard.edu/abs/2004ApJ...605..321S} {605, 321}

\bibitem[\protect\citeauthoryear{{Shi}, {Krolik}  \& {Hirose}}{{Shi}
  et~al.}{2010}]{Shi2010}
{Shi} J.,  {Krolik} J.~H.,   {Hirose} S.,  2010, \mn@doi [\apj]
  {10.1088/0004-637X/708/2/1716}, \href
  {http://adsabs.harvard.edu/abs/2010ApJ...708.1716S} {708, 1716}

\bibitem[\protect\citeauthoryear{{Simon} \& {Hawley}}{{Simon} \&
  {Hawley}}{2009}]{SH2009}
{Simon} J.~B.,  {Hawley} J.~F.,  2009, \mn@doi [\apj]
  {10.1088/0004-637X/707/1/833}, \href
  {http://adsabs.harvard.edu/abs/2009ApJ...707..833S} {707, 833}

\bibitem[\protect\citeauthoryear{{Simon}, {Hawley}  \& {Beckwith}}{{Simon}
  et~al.}{2009}]{simonetal2009}
{Simon} J.~B.,  {Hawley} J.~F.,   {Beckwith} K.,  2009, \mn@doi [\apj]
  {10.1088/0004-637X/690/1/974}, \href
  {http://adsabs.harvard.edu/abs/2009ApJ...690..974S} {690, 974}

\bibitem[\protect\citeauthoryear{{Simon}, {Hawley}  \& {Beckwith}}{{Simon}
  et~al.}{2011}]{SHB2011}
{Simon} J.~B.,  {Hawley} J.~F.,   {Beckwith} K.,  2011, \mn@doi [\apj]
  {10.1088/0004-637X/730/2/94}, \href
  {http://adsabs.harvard.edu/abs/2011ApJ...730...94S} {730, 94}

\bibitem[\protect\citeauthoryear{{Simon}, {Bai}, {Stone}, {Armitage}  \&
  {Beckwith}}{{Simon} et~al.}{2013a}]{Simon13a}
{Simon} J.~B.,  {Bai} X.-N.,  {Stone} J.~M.,  {Armitage} P.~J.,   {Beckwith}
  K.,  2013a, \mn@doi [\apj] {10.1088/0004-637X/764/1/66}, \href
  {http://adsabs.harvard.edu/abs/2013ApJ...764...66S} {764, 66}

\bibitem[\protect\citeauthoryear{{Simon}, {Bai}, {Armitage}, {Stone}  \&
  {Beckwith}}{{Simon} et~al.}{2013b}]{Simon13b}
{Simon} J.~B.,  {Bai} X.-N.,  {Armitage} P.~J.,  {Stone} J.~M.,   {Beckwith}
  K.,  2013b, \mn@doi [\apj] {10.1088/0004-637X/775/1/73}, \href
  {http://adsabs.harvard.edu/abs/2013ApJ...775...73S} {775, 73}

\bibitem[\protect\citeauthoryear{{Squire} \& {Bhattacharjee}}{{Squire} \&
  {Bhattacharjee}}{2015b}]{SB2015b}
{Squire} J.,  {Bhattacharjee} A.,  2015b, preprint, \href
  {http://adsabs.harvard.edu/abs/2015arXiv150801566S} {} (\mn@eprint {arXiv}
  {1508.01566})

\bibitem[\protect\citeauthoryear{{Squire} \& {Bhattacharjee}}{{Squire} \&
  {Bhattacharjee}}{2015a}]{squire2015a}
{Squire} J.,  {Bhattacharjee} A.,  2015a, preprint, \href
  {http://adsabs.harvard.edu/abs/2015arXiv150604109S} {} (\mn@eprint {arXiv}
  {1506.04109})

\bibitem[\protect\citeauthoryear{{Stone} \& {Gardiner}}{{Stone} \&
  {Gardiner}}{2010}]{sg10}
{Stone} J.~M.,  {Gardiner} T.~A.,  2010, \mn@doi [\apjs]
  {10.1088/0067-0049/189/1/142}, \href
  {http://adsabs.harvard.edu/abs/2010ApJS..189..142S} {189, 142}

\bibitem[\protect\citeauthoryear{{Stone}, {Hawley}, {Gammie}  \&
  {Balbus}}{{Stone} et~al.}{1996}]{shgb96}
{Stone} J.~M.,  {Hawley} J.~F.,  {Gammie} C.~F.,   {Balbus} S.~A.,  1996,
  \mn@doi [\apj] {10.1086/177280}, \href
  {http://adsabs.harvard.edu/abs/1996ApJ...463..656S} {463, 656}

\bibitem[\protect\citeauthoryear{{Stone}, {Gardiner}, {Teuben}, {Hawley}  \&
  {Simon}}{{Stone} et~al.}{2008}]{stoneetal08}
{Stone} J.~M.,  {Gardiner} T.~A.,  {Teuben} P.,  {Hawley} J.~F.,   {Simon}
  J.~B.,  2008, \mn@doi [\apjs] {10.1086/588755}, \href
  {http://adsabs.harvard.edu/abs/2008ApJS..178..137S} {178, 137}

\bibitem[\protect\citeauthoryear{{Suzuki} \& {Inutsuka}}{{Suzuki} \&
  {Inutsuka}}{2009}]{suzuki2009}
{Suzuki} T.~K.,  {Inutsuka} S.-i.,  2009, \mn@doi [\apjl]
  {10.1088/0004-637X/691/1/L49}, \href
  {http://adsabs.harvard.edu/abs/2009ApJ...691L..49S} {691, L49}

\bibitem[\protect\citeauthoryear{{Thaler} \& {Spruit}}{{Thaler} \&
  {Spruit}}{2015}]{TS2015}
{Thaler} I.,  {Spruit} H.~C.,  2015, \mn@doi [\aap]
  {10.1051/0004-6361/201423738}, \href
  {http://adsabs.harvard.edu/abs/2015A%26A...578A..54T} {578, A54}


\bibitem[\protect\citeauthoryear{{Turner}, {Sano}  \& {Dziourkevitch}}{{Turner}
  et~al.}{2007}]{tsd2007}
{Turner} N.~J.,  {Sano} T.,   {Dziourkevitch} N.,  2007, \mn@doi [\apj]
  {10.1086/512007}, \href {http://adsabs.harvard.edu/abs/2007ApJ...659..729T}
  {659, 729}

\bibitem[\protect\citeauthoryear{{Wardle}}{{Wardle}}{1999}]{Wardle1999}
{Wardle} M.,  1999, \mn@doi [\mnras] {10.1046/j.1365-8711.1999.02670.x}, \href
  {http://adsabs.harvard.edu/abs/1999MNRAS.307..849W} {307, 849}

\bibitem[\protect\citeauthoryear{{Winters}, {Balbus}  \& {Hawley}}{{Winters}
  et~al.}{2003}]{WBH2003}
{Winters} W.~F.,  {Balbus} S.~A.,   {Hawley} J.~F.,  2003, \mn@doi [\mnras]
  {10.1046/j.1365-8711.2003.06315.x}, \href
  {http://adsabs.harvard.edu/abs/2003MNRAS.340..519W} {340, 519}

\bibitem[\protect\citeauthoryear{{Yousef}, {Heinemann}, {Schekochihin},
  {Kleeorin}, {Rogachevskii}, {Iskakov}, {Cowley}  \& {McWilliams}}{{Yousef}
  et~al.}{2008}]{Yousefetal2008}
{Yousef} T.~A.,  {Heinemann} T.,  {Schekochihin} A.~A.,  {Kleeorin} N.,
  {Rogachevskii} I.,  {Iskakov} A.~B.,  {Cowley} S.~C.,   {McWilliams} J.~C.,
  2008, \mn@doi [Physical Review Letters] {10.1103/PhysRevLett.100.184501},
  \href {http://adsabs.harvard.edu/abs/2008PhRvL.100r4501Y} {100, 184501}

\bibitem[\protect\citeauthoryear{{Ziegler} \& {R{\"u}diger}}{{Ziegler} \&
  {R{\"u}diger}}{2000}]{ZR2000}
{Ziegler} U.,  {R{\"u}diger} G.,  2000, \aap, \href
  {http://adsabs.harvard.edu/abs/2000A%26A...356.1141Z} {356, 1141}


\makeatother
\end{thebibliography}

\bsp	

\label{lastpage}
\end{document}